\documentstyle[preprint,aps]{revtex}
\renewcommand{\theequation}{\arabic{section}.\arabic{equation}}
\begin{document}
\draft
\preprint{SNUTP 93-78}
\vskip 1cm
\title
{A Proposal of Positive-Definite Local Gravitational \\
Energy Density in General Relativity}

\author{J.H. Yoon\thanks{e-mail address: snu00162@krsnucc1.bitnet}}
\address{Center for Theoretical Physics, \\
Seoul National University, Seoul 151-742, Korea}

\date{\today}
\maketitle

\begin{abstract}
We propose a 4-dimensional Kaluza-Klein approach to general
relativity in the (2,2)-splitting of space-time using the
double null gauge. The associated Lagrangian density, implemented
with the auxiliary equations associated with the double null gauge,
is equivalent to the Einstein-Hilbert Lagrangian density, since
it yields the same field equations as the E-H Lagrangian density
does. It is describable as a (1+1)-dimensional Yang-Mills type gauge
theory coupled to (1+1)-dimensional matter fields,
where the minimal coupling associated with the
infinite dimensional diffeomorphism group  of the 2-dimensional
spacelike fibre space automatically appears. The physical
degrees of freedom of gravitational field show up as a
(1+1)-dimensional non-linear sigma model in our Lagrangian
density. Written in the first-order formalism, our Lagrangian
density directly yields a non-zero local Hamiltonian density,
where the associated time function is the retarded time.
{}From this Hamiltonian density, we obtain a positive-definite
local gravitational energy density. In the asymptotically flat
space-times, the volume integrals of the
proposed local gravitational energy density over
suitable 3-dimensional hypersurfaces correctly reproduce the Bondi
mass and the ADM mass expressed as surface integrals
at null and spatial infinity, respectively, supporting our
proposal. We also obtain the Bondi mass-loss formula as a
negative-definite flux integral of a bilinear in the
gravitational currents at null infinity.

\end{abstract}


\section{Introduction}
\label{sec:intro}

The exact correspondence of the Euclidean self-dual Einstein's
equations to the equations of motion of 2-dimensional non-linear
sigma models with the target space as the area-preserving
diffeomorphism of 2-surface\cite{park,parka} has inspired us
to look further into the intriguing question whether
the full-fledged general relativity of the 4-dimensional
space-time can be also formulated as a certain
(1+1)-dimensional field theory. Recently we have shown that
such a description is indeed possible, and constructed the
action principle\cite{soh,yoon,yoona,yoonb} in the
framework of the 4-dimensional Kaluza-Klein theory in the
(2,2)-splitting. In this approach, the 4-dimensional
space-time, at least for a finite range of space-time, is
viewed as a fibred manifold that consists of the
(1+1)-dimensional ``space-time" and the 2-dimensional
``auxiliary" fibre space.

There are certain advantages of this 4-dimensional KK approach
to general relativity in the above splitting, which led us to
develop this formalism. We list a few of them. First of all,
in (1+1)-dimensions, there exist a number of field
theoretic methods recently developed thanks to the
string-related theories. Hopefully the rich mathematical methods
in (1+1)-dimensions might also prove useful in studying
general relativity in the (2,2)-splitting, classical and quantum.
Moreover, in this KK formulation, general relativity can be
viewed as a (1+1)-dimensional {\it gauge} theory with the
prescribed interactions and auxiliary equations. Since
the major advantage of gauge theory formulation is that
gauge invariant quantities automatically solve
Gauss-law equations associated with the gauge invariance,
the problem of solving constraints of general relativity
could be made even {\it trivial},
at least for some of them. Furthermore, this formulation
allows us to {\it forget} about the space-time picture of
general relativity; instead, it enables us to study general
relativity much the way as we do for Yang-Mills theories
coupled to matter fields in (1+1)-dimensions, putting the
space-time physics into a new perspective.

This (1+1)-dimensional method, however, is not entirely
new since it was {\it virtually} used in analyzing gravitational
waves\cite{bondi,sach}, and was further developed
in the spin-coefficient
formalism\cite{newman,unti,newmantod,pen}
and the null hypersurface
formalism\cite{sacha,sachb,inv,inva,small,torr,hay}.
In these formalisms, a special gauge which we may call
the double null gauge is chosen such that two
real dual null vector fields whose congruences span the
(1+1)-dimensional submanifold are singled out,
and the Einstein's equations are spelled out in that gauge.
A characteristic feature of these formalisms, among others,
lies in that the true physical degrees of freedom of
gravitational field show up in the conformal 2-geometry of the
transverse 2-surface\cite{sach,sacha,sachb,inv,inva,small,adm}.
This feature, that has been particularly useful for
studying the propagation of gravitational waves
in the asymptotically flat space-times, further motivated
the canonical
analysis\cite{torr,gold,golda,goldb,kla,seif,naga}
of the null hypersurface formalism, in the hope of
getting quantum theory of gravity by quantizing the true physical
degrees of freedom of gravitational field.

In view of these advantages of the KK formalism and the
null hypersurface formalism, it therefore seems worth
combining both formalisms
to see what could be learnt more about general relativity.
In this article, we shall present such a formalism.
In this approach, it is the
(1+1)-dimensional submanifold spanned by two real dual null vector
fields that we imagine as ``space-time" and the remaining
transverse 2-surface as the ``auxiliary" fibre space.
As a by-product of our KK approach in the double null gauge,
we obtain a new result which we report in this article.
Namely, we propose a {\it positive-definite} {\it local}
gravitational energy density in general relativity {\it without}
referring to the
boundary conditions, and show that the volume integrals of the
proposed energy density over suitably chosen 3-dimensional
hypersurfaces correctly yield the
Bondi and the ADM surface integral at null and spatial infinity
in the asymptotically flat space-times, respectively.
We also obtain the Bondi
mass-loss formula as a negative-definite
flux integral of the gravitational degrees of freedom at
null infinity\cite{bondi,sach,newmantod}. The proposed local
gravitational energy density comes directly from the local
Hamiltonian density of general relativity described as the
4-dimensional KK theory in the double null gauge.
The associated time function is the {\it retarded} time,
and has the physical interpretation\cite{adma} as the
{\it phase} of the local gravitational radiation
in situations where gravitational waves are present.

This article is organized as follows. In section II, we present
the 4-dimensional KK theory in the (2,2)-splitting, using the
double null gauge.
It will be seen that, even when the gauge symmetry is an
infinite dimensional symmetry such as the group of
diffeomorphisms, the KK idea is still useful by showing that the
KK variables transform properly as gauge fields and tensor fields
under the corresponding gauge
transformations\cite{soh,yoon,yoona,yoonb,zoh}.
Next, we present the Lagrangian density for
general relativity in the double null gauge, which is implemented
by the auxiliary equations associated with the double null
gauge using the Lagrange multipliers. This Lagrangian density
is equivalent to the Einstein-Hilbert Lagrangian density, since it
yields the same field equations as the E-H Lagrangian density does.
We shall present the 10 Einstein's equations in the double
null gauge. Moreover, we shall find that 2 of these 10 equations
are in a form of the Schr\"{o}dinger equation,
reminiscent of the Brill wave equation\cite{brill,brilla,wheel},
i.e. the initial value equation of the axi-symmetric
gravitational waves at the moment of time symmetry.

In section III, we shall present this Lagrangian
density in the first-order formalism, with the retarded time
identified as our clock variable. This immediately leads to the
local Hamiltonian density and thus to the local gravitational
energy density that we are interested in. The proposed local
gravitational energy density is positive-definite.
We shall further show that the volume integral of the proposed
local gravitational energy density over a suitably chosen 3-dimensional
hypersurface become a surface integral, using the vacuum
Einstein's equations and the Bianchi identities.
In the asymptotically flat space-times, this surface
integral becomes the Bondi
and the ADM surface integral defined at null and
spatial infinity, respectively. We also derive the
Bondi mass-loss formula from the proposed gravitational energy
density.

In Appendix A,  we shall describe the general (2,2)-splitting
of space-time, and present the resulting E-H Lagrangian density
without picking up a special gauge, as we need it
when we wish to obtain the field equations from
the variational principle.
In Appendix B, we shall introduce the
covariant null tetrads, as we shall use them in Appendix C
where we show that the proposed volume integrals in section III
can be expressed as surface integrals.
In Appendix D, we shall show that $\kappa_{\pm}^{2}$, which
appears when we discuss the Bondi mass-loss, is positive-definite.

\section{The Lagrangian density in the Double Null Gauge}
\label{sec:kk}
\widetext
\setcounter{equation}{0}

In this section we combine the null hypersurface formalism
with the 4-dimensional Kaluza-Klein approach where space-time
is viewed as a fibred manifold, i.e. a local product of
the (1+1)-dimensional base manifold and the 2-dimensional
fibre space. Let the vector fields
\begin{math}
\partial / \! \partial X^{A}
=(\partial / \!  \partial u,\partial / \!
\partial v, \partial / \!  \partial y^{a})
\end{math} ($a=2,3$)
span the 4-dimensional space-time.
In a Lorentzian space-time we consider here,
there always exist two real null
vector fields, which we may choose orthogonal to the
2-dimensional spacelike surface $N_{2}$ spanned by
$\partial / \!  \partial y^{a}$.
Following the KK idea\cite{cho},
the two null vector fields can be represented as the linear
combinations of these basis vector fields
\begin{equation}
{\partial \over \partial u}-A_{+}^{\ a}
{\partial \over \partial y^{a}},
\hspace{.5cm} {\rm and} \hspace{.5cm}
{\partial \over \partial v}-A_{-}^{\ a}
{\partial \over \partial y^{a}},               \label{nullc}
\end{equation}
for some functions $A_{\pm}^{\ a}(u,v,y)$. Since these null vector
fields are assumed to be normal to $N_{2}$, the line element
may be written in a manifestly symmetric way as follows;
\begin{equation}
ds^{2} =- 2 du dv + \phi_{a b} (A_{+}^{\ a} du
       + A_{-}^{\ a} dv + dy^{a})
       (A_{+}^{\ b} du +A_{-}^{\ b} dv + dy^{b}),    \label{res}
\end{equation}
where $\phi_{a b}(u,v,y)$ is the 2-dimensional metric on $N_{2}$.
Notice that, as a consequence of picking up two null vector fields
normal to $N_{2}$, 2 out of the 10 metric
coefficient functions were gauged away in (\ref{res}).
In addition, one more function was removed from (\ref{res})
by choosing the coordinate $v$ such that
$C dv' = dv$ for some function $C$, i.e. by choosing $C=1$.
The elimination of these 3 functions may be viewed as a partial
gauge-fixing of the space-time diffeomorphism, and
may be better understood in terms of the dual metric,
which we may write
\begin{eqnarray}
& &g^{+ +}=g^{- -}=0, \hspace{.5cm}  g^{+ -}=g^{- +}=-1, \hspace{.5cm}
   g^{+ a}=A_{-}^{\ a},                         \nonumber\\
& &g^{- a}=A_{+}^{\ a}, \hspace{.5cm}
   g^{a b}=\phi^{a b} -2 A_{+}^{\ a}A_{-}^{\ b}.      \label{inv}
\end{eqnarray}
That $g^{+ +}=g^{- -}=0$ means that $du$ and $dv$ are dual null
vector fields, and that $g^{+ -}=-1$ is a normalization
condition for $v$, given an arbitrary function $u$.
We shall call this gauge as the double null
gauge\footnote{We notice that this double
null gauge is valid only for a finite range of space-time.
See for instance
\cite{sacha,sachb,brilla}.}, and general relativity formulated
in this gauge is referred to as the double null
formalism\cite{sacha,sachb,inv,inva,small}.
The 3-dimensional hypersurface defined by $u={\rm constant}$
is a null hypersurface since it is metrically degenerate;
within each null hypersurface the 2-dimensional spacelike
space $N_{2}$ defined by $v={\rm constant}$ is
transverse to both $du$ and $dv$.

In order to see whether this 4-dimensional KK program
is justifiable in the absence of any Killing symmetry,
as is the case here, we have to first examine the
transformation properties
of $\phi_{a b}$ and $A_{\pm}^{\ a}$ in (\ref{res})
under the action of some group of transformations associated
with $N_{2}$. The most natural group of transformations
associated with $N_{2}$ is the diffeomorphisms of $N_{2}$,
i.e. diff$N_{2}$. Under the diff$N_{2}$ transformation
\begin{equation}
y^{' a}=y^{' a} (u,v,y), \hspace{.5cm} u'=u,  \hspace{.5cm}
v'=v,                                 \label{coor}
\end{equation}
these fields must transform as
\renewcommand{\theequation}{\arabic{section}.5\alph{equation}}
\setcounter{equation}{0}
\begin{eqnarray}
& &\phi'_{a b}(u,v,y')={\partial y^{c} \over \partial y^{' a}}
   {\partial y^{d} \over \partial y^{' b}}
   \phi_{c d}(u,v,y),               \label{trane}\\
& &A_{\pm}^{\ 'a}(u,v,y')={\partial y^{' a} \over \partial y^{c}}
   A_{\pm}^{\ c}(u,v,y) -\partial_{\pm} y^{' a},   \label{tran}
\end{eqnarray}
\renewcommand{\theequation}{\arabic{section}.\arabic{equation}}
\setcounter{equation}{5}
\hspace{-.35cm}
so that the line element $ds^{2}$ remains
invariant\cite{soh,yoon,yoona,yoonb}.
Under the corresponding infinitesimal transformation
\begin{equation}
\delta y^{ a}=\xi^{a} (u,v,y), \hspace{.5cm}
\delta u=\delta v=0,
\end{equation}
where $\xi^{a}$ is an arbitrary function, we find
\renewcommand{\theequation}{\arabic{section}.7\alph{equation}}
\setcounter{equation}{0}
\begin{eqnarray}
& &\delta \phi_{a b}=-[\xi,  \phi]_{ a b},     \label{tranc}\\
& &\delta A_{\pm}^{\ a}=-D_{\pm}\xi^{a}
   =-\partial_{\pm}\xi^{a} + [A_{\pm}, \xi]^{a}, \label{trana}
\end{eqnarray}
\renewcommand{\theequation}{\arabic{section}.\arabic{equation}}
\setcounter{equation}{7}
\hspace{-.35cm}
where $A_{\pm}:=A_{\pm}^{\ a}\partial_{a}$ and
$\xi:=\xi^{a}\partial_{a}$. Here the brackets are
the Lie derivatives associated with diff$N_{2}$,
\renewcommand{\theequation}{\arabic{section}.8\alph{equation}}
\setcounter{equation}{0}
\begin{eqnarray}
& &[\xi,  \phi]_{ a b}=\xi^{c}\partial_{c}\phi_{a b}
   +(\partial_{a}\xi^{c})\phi_{c b}
   +(\partial_{b}\xi^{c})\phi_{a c},             \\
& &[A_{\pm}, \xi]^{a}=A_{\pm}^{\ c}\partial_{c}\xi^{a}
   - \xi^{c}\partial_{c}A_{\pm}^{\ a}.
\end{eqnarray}
\renewcommand{\theequation}{\arabic{section}.\arabic{equation}}
\setcounter{equation}{8}
\hspace{-.35cm}
This observation tells us two things. First, diff$N_{2}$ is
the residual symmetry\cite{newmantod} of the line
element (\ref{res}) which survives even after the double
null gauge was chosen.
Second, diff$N_{2}$ should be viewed as a
local\footnote{Here local means local in the
(1+1)-dimensional ``space-time".} gauge symmetry of
the Yang-Mills type, since $A_{\pm}^{\ a}$ and
$\phi_{a b}$ transform as a gauge field and a tensor field under
the diff$N_{2}$ transformations, respectively.
This feature is rather surprising, since the KK variables
were often thought to be useful for higher dimensional gravity
theories where the degrees of freedom
associated with the extra dimensions are suppressed in one
way or another\footnote{See however \cite{zoh,saund}.}.
In our 4-dimensional KK approach to general relativity,
we leave all the ``internal" degrees of freedom intact
so that all the fields in (\ref{res}) depend on all of the
coordinates $(u,v,y^{a})$.
In spite of these generalities, the Lagrangian density
associated with the metric (\ref{res})
can be still identified, \`{a} la Kaluza-Klein,
as a gauge theory Lagrangian density defined on the
(1+1)-dimensional ``space-time", with diff$N_{2}$ as the
associated local gauge symmetry, as we shall see shortly.

The metric (\ref{res}) in the double null gauge
can be obtained from the general KK line element
\begin{equation}
ds^{2}=\gamma_{\mu\nu}dx^{\mu}dx^{\nu}
+ \phi_{a b}(A_{\mu}^{\ a}dx^{\mu} + dy^{a})
            (A_{\nu}^{\ b}dx^{\nu} + dy^{b}),  \label{gene}
\end{equation}
where $\mu,\nu=0,1$ and $a,b=2,3$. From this
we obtain (\ref{res}) by introducing the retarded and advanced
coordinate $(u,v)$ and the fields $A_{\pm}^{\ a}$,
\renewcommand{\theequation}{\arabic{section}.10\alph{equation}}
\setcounter{equation}{0}
\begin{eqnarray}
& &u={1\over \sqrt{2}}(x^{0} - x^{1}), \hspace{.5cm}
v={1\over \sqrt{2}}(x^{0} + x^{1}),  \\
& &A_{\pm}^{\ a}={1\over \sqrt{2}}
   (A_{0}^{\ a} \mp A_{1}^{\ a}),        \label{nula}
\end{eqnarray}
\renewcommand{\theequation}{\arabic{section}.\arabic{equation}}
\setcounter{equation}{10}
\hspace{-.35cm}
assuming the (1+1)-dimensional ``space-time" metric
$\gamma_{\mu\nu}$ to be
\begin{equation}
\gamma_{+ -}=-1, \hspace{.5cm}
\gamma_{+ +}=\gamma_{- -}=0,       \label{gamma}
\end{equation}
in the $(u,v)$-coordinates, where $+(-)$ represents $u(v)$.
In Appendix A, the general E-H Lagrangian
density\cite{soh,yoon,yoona} for the metric (\ref{gene})
and the prescription how to obtain it are
presented.
Using the ``ansatz" (\ref{gamma}), we can easily show
that the general E-H Lagrangian density (\ref{ac})
reduces to the following expression
corresponding to the metric (\ref{res}).
If we neglect the auxiliary equations associated
with the double null gauge (\ref{gamma}) for the moment,
it is given by\cite{yoona}
\begin{equation}
{\cal L}_{0}=\sqrt{\phi} \, \Big[
    {1\over 2}\phi_{a b}F_{+-} ^ { \  \ a}F_{+-}^{\ \ b}
   +{1\over 2}\phi ^ {a b}\phi ^ {c d}
   \Big\{
   (D_{+}\phi_{a c})(D_{-}\phi_{b d})
  -(D_{+}\phi_{a b})(D_{-}\phi_{c d})\Big\} \Big], \label{ga}
\end{equation}
where we ignored the surface terms. Here
\begin{math}
\phi={\rm det}\, \phi_{a b},
\end{math}
and
$F_{+-}^{\ \ a}$ is the diff$N_{2}$-valued field strength,
and $D_{\pm}\phi_{a b}$
is the diff$N_{2}$-covariant derivative defined as
\renewcommand{\theequation}{\arabic{section}.13\alph{equation}}
\setcounter{equation}{0}
\begin{eqnarray}
& &F_{+-}^{\ \ a}=\partial_{+} A_{-} ^ { \ a}-\partial_{-}
  A_{+} ^ { \ a} - [A_{+}, A_{-}]^{a},       \label{need}  \\
& &D_{\pm}\phi_{a b} = \partial_{\pm}\phi_{a b}
   - [{A_{\pm}}, \phi]_{ a b},          \label{fie}
\end{eqnarray}
\renewcommand{\theequation}{\arabic{section}.\arabic{equation}}
\setcounter{equation}{13}
\hspace{-.35cm}
where $[A_{+}, A_{-}]^{a}$ and $[{A_{\pm}}, \phi]_{ a b}$
are the Lie derivatives defined as
\renewcommand{\theequation}{\arabic{section}.14\alph{equation}}
\setcounter{equation}{0}
\begin{eqnarray}
& &[A_{+}, A_{-}]^{a}=A_{+}^{\ c}\partial_{c}A_{-}^{\ a}
   -A_{-}^{\ c}\partial_{c}A_{+}^{\ a},   \\
& &[{A_{\pm}}, \phi]_{ a b}=A_{\pm} ^ { \ c}\partial_c \phi_{a b}
    +(\partial_a A_{\pm} ^ {\ c})\phi_{c b}
    +(\partial_b A_{\pm} ^ { \ c})\phi_{a c},  \label{cov}
\end{eqnarray}
\renewcommand{\theequation}{\arabic{section}.\arabic{equation}}
\setcounter{equation}{14}
\hspace{-.35cm}
respectively (see Appendix A).
Recall that, in the double null formalism of general relativity,
the transverse 2-metric with a unit determinant,
i.e. the conformal 2-geometry, is the two physical degrees of
freedom of gravitational
field\cite{sach,sacha,sachb,inv,inva,small}.
Since we are essentially reformulating
the double null formalism from the KK point of view in
this article, we would like to see first of all what the
Lagrangian density looks like when written in terms of the
conformal 2-geometry. Let us therefore
decompose the 2-metric $\phi_{a b}$ into the conformal classes
\begin{equation}
\phi_{a b}=\Omega\rho_{a b}, \hspace{.5cm}
(\Omega > 0 \ \  \  {\rm and} \ \ \
{\rm det}\, \rho_{a b} = 1),                \label{dec}
\end{equation}
where $\rho_{a b}$
is the conformal 2-geometry\footnote{This may be viewed as
a finite analogue of the physical transverse traceless degrees
of freedom of the spin-2 fields propagating in the flat
space-time\cite{sachb}.
The double null gauge may be also viewed as an
analogue of the Coulomb gauge in Maxwell's theory, where
the physical degrees of freedom are the transverse traceless
vector potentials\cite{komar}.}
of the transverse 2-surface $N_{2}$.
If we define $\sigma$ by $\sigma:={\rm ln}\, \Omega$,
the second term in (\ref{ga}) becomes
\begin{eqnarray}
K&:=&{1\over 2}\sqrt{\phi} \, \phi ^ {a b}\phi ^ {c d}
         \Big\{
        (D_{+}\phi_{a c})(D_{-}\phi_{b d})
        -(D_{+}\phi_{a b})(D_{-}\phi_{c d})\Big\}   \nonumber\\
&=&-\Omega^{-1}(D_{+} \Omega)(D_{-} \Omega)
  +{1\over 2}\Omega \rho^{a b}\rho^{c d}
    (D_{+}\rho_{a c})(D_{-}\rho_{b d})      \nonumber\\
&=&-{\rm e}^{\sigma}(D_{+}\sigma) (D_{-}\sigma)
   +{1\over 2}{\rm e}^{\sigma}\rho^{a b}\rho^{c d}
    (D_{+}\rho_{a c})(D_{-}\rho_{b d}),       \label{kin}
\end{eqnarray}
where $D_{\pm}\Omega$, $D_{\pm}\sigma$, and $D_{\pm}\rho_{a b}$
are the diff$N_{2}$-covariant derivatives
\renewcommand{\theequation}{\arabic{section}.17\alph{equation}}
\setcounter{equation}{0}
\begin{eqnarray}
& &D_{\pm}\Omega = \partial_{\pm}\Omega - [A_{\pm}, \Omega],  \\
& &D_{\pm}\sigma=\partial_{\pm}\sigma
  - [A_{\pm}, \sigma],   \label{den}\\
& &D_{\pm}\rho_{a b}=\partial_{\pm}\rho_{a b}
   - [A_{\pm}, \rho]_{a b},            \label{rho}
\end{eqnarray}
\renewcommand{\theequation}{\arabic{section}.\arabic{equation}}
\setcounter{equation}{17}
\hspace{-.35cm}
and $[A_{\pm}, \Omega]$,  $[A_{\pm}, \sigma]$, and
$[A_{\pm}, \rho]_{a b}$ are given by
\renewcommand{\theequation}{\arabic{section}.18\alph{equation}}
\setcounter{equation}{0}
\begin{eqnarray}
& &[A_{\pm}, \Omega]=A_{\pm}^{\ a}\partial_{a}\Omega
    +(\partial_{a}A_{\pm}^{\ a}) \Omega,          \\
& &[A_{\pm}, \sigma]=A_{\pm}^{\ a}\partial_{a}\sigma
    +\partial_{a}A_{\pm}^{\ a},                 \\
& &[{A_{\pm}}, \rho]_{a b}=
    A_{\pm} ^ { \ c}\partial_c \rho_{a b}
    +(\partial_a A_{\pm} ^ { \ c})\rho_{c b}
    +(\partial_b A_{\pm} ^ { \ c})\rho_{a c}
    -(\partial_c A_{\pm} ^ { \ c})\rho_{a b},      \label{rhoa}
\end{eqnarray}
\renewcommand{\theequation}{\arabic{section}.\arabic{equation}}
\setcounter{equation}{18}
\hspace{-.35cm}
respectively. Here $\partial_{a}A_{\mu}^{\ a}$-terms
are included in the Lie derivatives,
since $\Omega$ and $\rho_{a b}$ are tensor
densities of weight $-1$ and $+1$ under diff$N_{2}$,
respectively. Thus (\ref{ga}) becomes
\begin{equation}
{\cal L}_{0}={1\over 2}{\rm e}^{2\sigma}\rho_{a b}
  F_{+-}^{\ \ a}F_{+-}^{\ \ b}
 -{\rm e}^{\sigma}(D_{+}\sigma) (D_{-}\sigma)
 +{1\over 2}{\rm e}^{\sigma}\rho^{a b}\rho^{c d}
 (D_{+}\rho_{a c})(D_{-}\rho_{b d}).      \label{lag}
\end{equation}

In order to find the correct variational principle
that yields all the 10 Einstein's equations, however, we must
implement (\ref{lag}) with the auxiliary equations associated
with the double null gauge. These equations can be found by first
varying the general E-H Lagrangian density (\ref{ac}) with
respect to $\gamma^{++}$,$\gamma^{--}$, and
$\gamma^{+-}$,
and then plugging the double null gauge (\ref{gamma})
into the resulting equations. They are found to be
\renewcommand{\theequation}{\arabic{section}.20\alph{equation}}
\setcounter{equation}{0}
\begin{eqnarray}
C_{\pm}&:=&D_{\pm}^{2}\sigma + {1\over 2}(D_{\pm}\sigma)^{2}
 + {1\over 4}\rho^{a b}\rho^{c d} (D_{\pm}\rho_{a c})
   (D_{\pm}\rho_{b d})=0,                 \label{cpm}\\
C_{0}&:=&-{1\over 2}{\rm e}^{\sigma}\rho_{a b}
  F_{+-}^{\ \ a}F_{+-}^{\ \ b}
 +D_{+}D_{-}\sigma + D_{-}D_{+}\sigma + 2(D_{+}\sigma)
  (D_{-}\sigma) +R_{2}             \nonumber\\
&=&0,                                \label{co}
\end{eqnarray}
\renewcommand{\theequation}{\arabic{section}.\arabic{equation}}
\setcounter{equation}{20}
\hspace{-.35cm}
respectively, where $R_{2}:=\phi^{a c}R_{a c}$ is the scalar
curvature of $N_{2}$. Remarkably, the equations
$C_{\pm}=0$ in (\ref{cpm}) may be written in a form
of the Schr\"{o}dinger equation. Using the identity
\begin{equation}
{\rm e}^{ \sigma /  2}\Big(D_{\pm}^{2}\sigma
 + {1\over 2}(D_{\pm}\sigma)^{2}\Big)
 =2\ D_{\pm}^{2}{\rm e}^{ \sigma / 2},         \label{sig}
\end{equation}
the equations $C_{\pm}=0$ become
\begin{equation}
D_{\pm}^{2}{\rm e}^{ \sigma / 2}+ \kappa_{\pm}^{2}
   {\rm e}^{ \sigma / 2}=0,  \hspace{.5cm}
{\rm where}\hspace{.5cm}
\kappa_{\pm}^{2}:={1\over 8}\rho^{a b}\rho^{c d}
(D_{\pm}\rho_{a c})  (D_{\pm}\rho_{b d}) \geq 0.  \label{geq}
\end{equation}
That $\kappa_{\pm}^{2}$, a bilinear in the {\it currents} of
the gravitational degrees of freedom, is positive-definite
can be shown easily (see Appendix D).
The equations (\ref{geq}) are the analogues of
the Brill wave equation\footnote{Our Schr\"{o}dinger
equations look like
one-dimensional wave equations coupled to gauge fields,
but actually they are 3-dimensional partial differential
equations like the Brill wave equation, due to
the Lie derivatives along $A_{\pm}=A_{\pm}^{\ a}\partial_{a}$
in (\ref{geq}).
The wave function in (\ref{geq}) is
related to the conformal factor of the 2-dimensional
wavefront, the spatial projection of the null
hypersurface $u={\rm constant}$, rather than that of spacelike
hypersurface. But it should be also mentioned that,
for the metric (\ref{res}), the area measure of the 2-dimensional
wavefront and the volume measure of 3-dimensional null
hypersurface $u={\rm constant}$ are the
same\cite{wald}.}\cite{brill,brilla,wheel},
as they are of the Schr\"{o}dinger equation type
for a wave function corresponding to a state of
zero energy in the potential $-\kappa_{\pm}^{2}$,
coupled to the external gauge fields $A_{\pm}^{\ a}$!
Thus, viewed as a scattering problem, the scattering
data $e^{\sigma / 2}$ in (\ref{geq}) is an auxiliary field
that can be determined by the potential $-\kappa_{\pm}^{2}$ up to
some integral ``constant" functions.
The generic behaviors of solutions
of the 2 Einstein's equations $C_{\pm}=0$ are therefore
expected either of
the scattering type, or of the bound-state or resonance type,
corresponding to the asymptotically flat space-times or
spatially closed universes, respectively,
on a par with the Brill wave equation.

The correct variational principle is now given by
\begin{equation}
{\cal L}={1\over 2}{\rm e}^{2\sigma}\rho_{a b}
  F_{+-}^{\ \ a}F_{+-}^{\ \ b}
 -{\rm e}^{\sigma}(D_{+}\sigma) (D_{-}\sigma)
 +{1\over 2}{\rm e}^{\sigma}\rho^{a b}\rho^{c d}
 (D_{+}\rho_{a c})(D_{-}\rho_{b d})
 + \sum_{\alpha = \pm, 0}
  \lambda^{\alpha}C_{\alpha},                 \label{finala}
\end{equation}
where $\lambda^{\alpha}$'s are the Lagrange multipliers
which should be put to zero after variation.
The equations of motions for $A_{\pm}^{\ a}$, $\sigma$, and
$\rho_{a b}$ (subject to ${\rm det} \, \rho_{a b}=1$) can be
obtained by varying (\ref{finala}), with $\lambda^{\alpha}=0$.
Here we present the results only;
\renewcommand{\theequation}{\arabic{section}.24\alph{equation}}
\setcounter{equation}{0}
\begin{eqnarray}
&(a)\ & D_{-}\Big( {\rm e}^{2\sigma}\rho_{a b}F_{+-}^{\ \ b}\Big)
    + {\rm e}^{\sigma}(D_{-}\sigma)
    (\partial_{a}\sigma)
    - \partial_{a} ({\rm e}^{\sigma} D_{-}\sigma)
    - {1\over 2}{\rm e}^{\sigma}\rho^{b c}\rho^{d e}
    (D_{-}\rho_{b d})(\partial_{a}\rho_{c e})      \nonumber\\
& \ & + \partial_{b}
\Big(
{\rm e}^{ \sigma}\rho^{b c}D_{-}\rho_{a c} \Big)=0,\label{eqq}\\
&(b)\ & D_{+}\Big( {\rm e}^{2\sigma}\rho_{a b}F_{+-}^{\ \ b}\Big)
    - {\rm e}^{\sigma}(D_{+}\sigma)
    (\partial_{a}\sigma)
   + \partial_{a} ({\rm e}^{\sigma} D_{+}\sigma)
   +{1\over 2}{\rm e}^{\sigma}\rho^{b c}\rho^{d e}
   (D_{+}\rho_{b d})(\partial_{a}\rho_{c e})         \nonumber\\
& \ &-\partial_{b}
 \Big(
{\rm e}^{ \sigma}\rho^{b c}D_{+}\rho_{a c} \Big)=0,\label{eqa}\\
&(c)\  & (D_{+}\sigma)(D_{-}\sigma) + 2 D_{{(}+}D_{-{)}}\sigma
    +{1\over 2}\rho^{a b}\rho^{c d}
    (D_{+}\rho_{a c})(D_{-}\rho_{b d}) +
{\rm e}^{\sigma}\rho_{a b}F_{+-}^{\ \ a}F_{+-}^{\ \ b} \nonumber\\
& \ & =0, \                                        \label{eqb}\\
&(d)\ &  D_{{(}+}\Big( {\rm e}^{\sigma}
     \rho^{a c}D_{-{)}}\rho_{b c} \Big)
    - {1\over 2}{\rm e}^{2 \sigma} \Big(
     \rho_{b c} F_{+-}^{\ \ a}F_{+-}^{\ \ c}
    -{1\over 2}\delta^{a}_{\ b}\rho_{c d}
     F_{+-}^{\ \ c}F_{+-}^{\ \ d} \Big) = 0, \     \label{eqc}
\end{eqnarray}
\renewcommand{\theequation}{\arabic{section}.\arabic{equation}}
\setcounter{equation}{24}
\hspace{-.35cm}
where the symmetric symbol is normalized such that
$(\alpha\beta):=(\alpha\beta + \beta\alpha)/  2$.
Together with the 3 equations $C_{\pm}=0$, $C_{0}=0$
that we obtain by varying (\ref{finala}) with respect to
$\lambda^{\pm}$, $\lambda^{0}$,
these field equations are identical to the
10 Einstein's equations spelled out in the double null gauge,
which we obtain by first varying the general E-H
Lagrangian density (\ref{ac}) in Appendix A, and then imposing
the double null gauge (\ref{gamma}).
Therefore the
Lagrangian density (\ref{finala}) is equivalent to the
general E-H Lagrangian density (\ref{ac}),
with the understanding that $\lambda^{\alpha}$'s are
to be set to zero after variation.

The Lagrangian density (\ref{finala}) may be
naturally interpreted as
the Yang-Mills type Lagrangian density on the (1+1)-dimensional
``space-time", interacting with the (1+1)-dimensional ``matter"
fields $\sigma$ and $\rho_{a b}$. The corresponding local
gauge symmetry is the {\it built-in} diff$N_{2}$,
and the ``matter" fields couple to the diff$N_{2}$-valued
gauge fields through the minimal couplings. In addition,
each term in (\ref{finala}),
including the auxiliary equations, is manifestly invariant
under the diff$N_{2}$ transformations. Therefore (\ref{finala})
should be duly regarded as a gauge theory formulation of
the vacuum general relativity\cite{yoon,yoona,yoonb}.

That $\rho_{a b}$ is the physical degrees of freedom can be also
seen as follows. In this (1+1)-dimensional interpretation,
the diff$N_{2}$-valued gauge fields $A_{\pm}^{\ a}$ are auxiliary
fields since they have no propagating (i.e. no transverse
traceless) degrees of freedom. Moreover, as we have seen already,
$\sigma$ is also an auxiliary field that
is determined by $\rho_{a b}$ through the equations
$C_{\pm}=0$. This confirms that
the two physical degrees of freedom of gravitational field
are indeed contained in $\rho_{a b}$. It seems appropriate to
notice here that, in the propagating equations of motion (\ref{eqc})
for $\rho_{a b}$, the source term is given by
\begin{equation}
{1\over 2}{\rm e}^{2 \sigma}\rho_{b c}
F_{+-}^{\ \ a}F_{+-}^{\ \ c},
\end{equation}
whose trace is precisely the local gravitational energy density,
as we shall see in the next section. This indicates that the
local energy density in general relativity indeed plays the
analogous role as the local charge density does in Maxwell's theory.

\section{The Local Gravitational Energy Density}
\label{sec:ham}
\setcounter{equation}{0}

In this section, we shall find the local Hamiltonian density of
general relativity. This can be obtained simply by
writing the local Lagrangian density (\ref{finala}) in the
first-order form using a suitable time coordinate.
The most natural time in this formulation seems the
retarded time $u$\cite{newmantod,adma}.
With the retarded time as our clock, the first term
in (\ref{finala}) may be written as
\begin{eqnarray}
{\cal L}_{\rm YM}&:=&{1\over 2}{\rm e}^{2\sigma}\rho_{a b}
   F_{+-}^{\ \ a}F_{+-}^{\ \ b}           \nonumber\\
&=& {\rm e}^{2\sigma}\rho_{a b}F_{+-}^{\ \ b}
    \Big( \partial_{+}A_{-}^{\ a} - \partial_{-}A_{+}^{\ a}
    -[ A_{+},  A_{-} ]^{a} - {1\over 2}F_{+-}^{\ \ a}
     \Big).                             \label{laaa}
\end{eqnarray}
In terms of the phase space variables $(\Pi_{a}, A_{-}^{\ a})$,
where $\Pi_{a}$ is defined as
\begin{equation}
\Pi_{a}={\rm e}^{2 \sigma}\rho_{a b}
   F_{+-}^{\ \ b},                     \label{mom}
\end{equation}
this can be written as
\begin{equation}
{\cal L}_{\rm YM}=\Pi_{a}\partial_{+} A_{-}^{\ a}
  -{1\over 2}{\rm e}^{-2 \sigma}\rho^{a b}\Pi_{a}\Pi_{b}
  + A_{+}^{\ a} D_{-}\Pi_{a},                   \label{ym}
\end{equation}
ignoring the surface terms. Here $D_{-}\Pi_{a}$
is the diff$N_{2}$-covariant derivative of the density $\Pi_{a}$
defined as
\begin{equation}
D_{-}\Pi_{a}=\partial_{-}\Pi_{a}-[A_{-}, \Pi ]_{ a}, \label{dmom}
\end{equation}
where $[A_{-}, \Pi ]_{ a}$ is the Lie derivative of $\Pi_{a}$,
\begin{equation}
[A_{-}, \Pi ]_{a}=A_{-}^{ \ c}\partial_{c}\Pi_{a}
   + (\partial_{a}A_{-}^{\ c})\Pi_{c} +
   (\partial_{c}A_{-}^{\ c})\Pi_{a}.              \label{first}
\end{equation}
The second and third term in (\ref{finala}) are already in
the first-order form, apart from the terms proportional to
$A_{+}^{\ a}$ whose variation yields the Gauss-law equations
associated with the residual diff$N_{2}$ invariance.
Putting these all together, the Lagrangian density (\ref{finala})
can be written in the following Hamiltonian
form\footnote{Notice that the Hessian of this Lagrangian
density is zero, so that the usual method of
Hamiltonization does not work. This is the reason that we did
not introduce the momenta conjugate to
$\sigma$ and $\rho_{a b}$.
See for instance\cite{wood,ish,git,fad}.}
\begin{eqnarray}
{\cal L}& = &  \Pi_{a}\partial_{+} A_{-}^{\ a}
 -{\rm e}^{ \sigma}(D_{-}\sigma) (\partial_{+}\sigma )
 + {1\over 2}{\rm e}^{ \sigma}
   \rho^{a b}\rho^{c d}(D_{-}\rho_{b d})
   (\partial_{+}\rho_{a c})      \nonumber\\
& & -{1\over 2}{\rm e}^{-2 \sigma}\rho^{a b}\Pi_{a}\Pi_{b}
+A_{+}^{\ a}C_{a}
+\sum_{\alpha=\pm, 0}\lambda^{\alpha}C_{\alpha},  \label{firsta}
\end{eqnarray}
where $C_{a}$ is given by
\begin{eqnarray}
C_{a}&=& D_{-}\Pi_{a}+{\rm e}^{\sigma}(D_{-}\sigma)
    (\partial_{a} \sigma)
   - \partial_{a}({\rm e}^{\sigma}D_{-}\sigma)
   -{1\over 2}{\rm e}^{\sigma}
    \rho^{b c}\rho^{d e}(D_{-}\rho_{b d})
    (\partial_{a}\rho_{c e})                        \nonumber\\
& &   +\partial_{b} \Big(  {\rm e}^{\sigma}
  \rho^{b c}D_{-}\rho_{a c}  \Big),                 \label{gaus}
\end{eqnarray}
which is the same as (\ref{eqq}) if we use (\ref{mom}).
{}From (\ref{firsta}) the
local gravitational Hamiltonian density $\cal H$ is given by
\begin{equation}
{\cal H}= {1\over 2}{\rm e}^{-2 \sigma}\rho^{a b}
\Pi_{a}\Pi_{b} - A_{+}^{\ a}C_{a} -
\sum_{\alpha = \pm, 0}\lambda^{\alpha}C_{\alpha}.   \label{ham}
\end{equation}
Using the 5 equations $C_{a}=0$, $C_{\pm}=0$, and $C_{0}=0$,
the local Hamiltonian density (\ref{ham}) becomes
\begin{equation}
{\cal E}={1\over 2}{\rm e}^{-2 \sigma}\rho^{a b}
\Pi_{a}\Pi_{b}
={1 \over 2}{\rm e}^{2 \sigma}\rho_{a b}
F_{+ -}^{\ \ a}F_{+ -}^{\ \ b}
\geq 0,          \label{ener}
\end{equation}
which is positive-definite for {\it any} $\sigma$ and $\Pi_{a}$,
since the conformal 2-metric $\rho_{a b}$ has a positive-definite
signature.
Thus, at least formally, we have obtained, in the double null
gauge, a positive-definite local gravitational energy
density for the vacuum general
relativity! The time function associated with this non-zero local
energy density is the retarded time $u$ that we
may choose at will\footnote{For the asymptotically flat
space-times, the number of possible choices of the
retarded time $u$ is equal to the number of an arbitrary,
monotonically increasing function of three variables
$(u, y^{a})$\cite{newmantod}. This may be true for
other space-times as well.}.
This is our proposal of the positive-definite
local gravitational energy density in this
article. This seems to be against the usual argument
that, in general relativity, local gravitational energies
can not be defined because they can be always ``transformed"
away due to the equivalence principle, let alone
the positive-definiteness. In our definition of the
local gravitational energy density, however,
the ``field strength" $F_{+ -}^{ \ \ a}$ in (\ref{need}) is
the coefficient of the commutator
of the two {\it null} vector fields
\begin{math}
\partial_{\pm}-A_{\pm}^{\ a}\partial_{a},
\end{math}
which measures the {\it twist} of the parallelogram
made of two successive parallel
transports of these null vector fields along each other.
Certainly the twist of this {\it null} parallelogram
can {\it not} be ``transformed" away even in a local Lorentz frame,
and thus can serve as {\it a} measure of gravitational energy
associated with the parallelogram
surrounding the space-time point under
consideration. The proposed local gravitational
energy density is just the square of this local ``field strength"
multiplied by the canonical integration measure.

In order to appreciate what this really means, however,
we have to first define the volume integral of the local
gravitational energy
density ${\cal E}$ over a 3-dimensional hypersurface
defined by $u={\rm constant}$, and evaluate it for the
asymptotically flat space-times, since the total gravitational
energy is well-defined only for the asymptotically flat
space-times. As we now show, for the asymptotically flat
space-times, the volume integrals of the proposed local
energy density over suitably chosen 3-dimensional
hypersurfaces can be re-expressed as the Bondi and ADM
surface integral at null and spatial infinity,
respectively. We shall also derive
the Bondi mass loss-formula as
a negative-definite flux integral of a bilinear in the
gravitational currents at null infinity.

\subsection{The Bondi Mass}
\label{subsec:bondi}

In this subsection we wish to show that, for the
asymptotically flat space-times, the volume integral of
(\ref{ener}) over the $u={\rm constant}$ null
hypersurface is precisely the Bondi mass as measured at
null infinity. Let us first notice that the volume integral $E$,
where
\begin{equation}
E={1\over 2}\int dv d^{2}y \ {\rm e}^{2\sigma}\rho_{a b}
     F_{+-}^{\ \ a}F_{+-}^{\ \ b} \geq 0,     \label{totalham}
\end{equation}
is positive-definite for {\it any}
topology of $N_{2}$\cite{unti}.
In order to express (\ref{totalham}) as a surface
integral, it is necessary to write it
in a slightly different form using the field equations.
For this let us consider the following identity
\begin{equation}
D_{+}D_{-}\sigma - D_{-}D_{+}\sigma
= - F_{+-}^{\ \ a}\partial_{a}\sigma
- \partial_{a}F_{+-}^{\ \ a}.                     \label{ddd}
\end{equation}
Using (\ref{ddd}), the integral of the equation $C_{0}=0$
in (\ref{co}) over
$N_{2}$ with the integration measure
${\rm e}^{\sigma}$ may be written as
\begin{equation}
{1\over 2}\int d^{2}y \ {\rm e}^{2\sigma}\rho_{a b}
F_{+-}^{\ \ a}F_{+-}^{\ \ b}
=\int d^{2}y \ {\rm e}^{\sigma}\Big\{ R_{2}
 + 2 (D_{+}\sigma)(D_{-}\sigma)
 + 2 D_{+}D_{-}\sigma   \Big\}
 +\int d^{2}y \ \partial_{a}
 ({\rm e}^{\sigma}F_{+-}^{\ \ a}).             \label{auxa}
\end{equation}
The last term in (\ref{auxa}) is zero
for any 2-surface $N_{2}$ that we assume compact without boundary.
Thus (\ref{totalham}) becomes
\begin{equation}
E=\int dv  d^{2} y \ {\rm e}^{\sigma} \Big\{
  R_{2}+ 2(D_{+}\sigma)(D_{-}\sigma)
   + 2 D_{+}D_{-}\sigma  \Big\}.         \label{ha1}
\end{equation}
Let us also integrate the equation (\ref{eqb}) over $N_{2}$,
using (\ref{ddd}), to obtain
\begin{eqnarray}
& &\int d^{2}y \ {\rm e}^{\sigma}
   \Big\{ (D_{+}\sigma)(D_{-}\sigma)
  + 2 D_{+}D_{-}\sigma \Big\}              \nonumber\\
& &=-\int d^{2}y \ {\rm e}^{2\sigma}\rho_{a b}
   F_{+-}^{\ \ a}F_{+-}^{\ \ b}
-{1\over 2}\int d^{2}y \ {\rm e}^{\sigma}\rho^{a b}\rho^{c d}
   (D_{+}\rho_{a c})(D_{-}\rho_{b d}) \nonumber\\
& &-\int d^{2}y \ \partial_{a}
  ({\rm e}^{\sigma}F_{+-}^{\ \ a}),                \label{int}
\end{eqnarray}
where the last term may be also dropped.
Thus the volume integral (\ref{ha1}) becomes
\begin{eqnarray}
E&=&\int dv d^{2} y \ {\rm e}^{\sigma} \Big\{  R_{2}
  + (D_{+}\sigma)(D_{-}\sigma)
  - {1\over 2}\rho^{a b}\rho^{c d}(D_{+}\rho_{a c})
   (D_{-}\rho_{b d}) \Big\}                        \nonumber\\
 & & -\int dv d^{2}y \ {\rm e}^{2\sigma} \rho_{a b}
   F_{+-}^{\ \ a}F_{+-}^{\ \ b},                \label{ha2}
\end{eqnarray}
or,
\begin{eqnarray}
E&=&{1\over 2}\int dv d^{2} y \ {\rm e}^{2\sigma}\rho_{a b}
     F_{+-}^{\ \ a}F_{+-}^{\ \ b}       \nonumber\\
 &=&{1\over 3}\int dv d^{2} y \ {\rm e}^{\sigma} \Big\{  R_{2}
  + (D_{+}\sigma)(D_{-}\sigma)
  - {1\over 2}\rho^{a b}\rho^{c d}(D_{+}\rho_{a c})
  (D_{-}\rho_{b d})\Big\}.                       \label{ha3}
\end{eqnarray}
To show that this\footnote{Notice that
this energy integral is different
from the one in Hayward's paper\cite{haya}.
The energy density he proposed is the minus of
${\cal L}_{0}$ in (\ref{lag}),
modulo the Euler density, and is
not positive-definite. This difference may be traced back to the
fact that in his paper both $u$ and $v$ are treated as the
time variables.}
can be expressed as a surface integral,
the covariant null tetrad notation
that we described in Appendix B is useful.
Let us notice that the Gauss equation
in the (2,2)-splitting of space-time is
given by\footnote{Here we used the vacuum Einstein's equations.}
\cite{haya,ash,asha,ashb}
\begin{equation}
 R_{2}  + (D_{+}\sigma)(D_{-}\sigma)
 - {1\over 2}\rho^{a b}\rho^{c d}
 (D_{+}\rho_{a c})(D_{-}\rho_{b d})
 =h^{A C}h^{B D} C_{A B C D},                 \label{gauss}
\end{equation}
where $C_{A B C D} (A,B,\cdots = 0,1,2,3)$ is
the conformal curvature tensor,
and $h_{A B}$ is the 2-metric on the transverse
surface $N_{2}$, i.e. the covariant form of $\phi_{a b}$.
Then the volume integral $E$ may be written as
\begin{equation}
E={1\over 3}\int dv d^{2}y \ {\rm e}^{\sigma}
h^{A C}h^{B D} C_{A B C D}.            \label{ha4}
\end{equation}
In Appendix C, we have shown that, using the
Bianchi identity
\begin{equation}
\nabla_{{[}M}C_{A B{]} C D}=0,     \label{bia}
\end{equation}
this can be expressed as the surface integral
\begin{equation}
E={1\over 3}\int dv d^{2}y \ {\rm e}^{\sigma}
   h^{A C}h^{B D}C_{A B C D}
   ={1\over 3}\lim \ v \!\! \int d^{2}y \ {\rm e}^{\sigma}
   h^{A C}h^{B D}C_{A B C D},                \label{bmass}
\end{equation}
where $\lim$ means that the integral over $N_{2}$ is to be evaluated
at the limiting boundary value(s) of $v$. This expression
picks up the coefficient of $1/v$-term, and
becomes precisely the Bondi mass\footnote{We have
a factor of $1/ 3$, which could be taken care of by a suitable
normalization.}\cite{haya,ash,asha,ashb} in the limit as
$v$ approaches to infinity!
Notice that the parameter $v/ \sqrt{2}$ becomes the
area radius in the limit $v\rightarrow \infty$ (keeping
$u=u_{0}={\rm constant}$), as our metric (\ref{res}) approaches to
\begin{equation}
ds^{2} \rightarrow -2dudv + {1\over 2}(v-u)^{2}(d\vartheta^{2}
+ {\rm sin}^{2}\vartheta d\varphi^{2})        \label{null}
\end{equation}
at null infinity.

\subsection{The Bondi Mass-Loss Formula}
\label{subsec:loss}

In this subsection we shall continue to obtain the Bondi
mass-loss formula in the presence of
the gravitational radiation in the asymptotically flat
space-times. For this, we may simply take a $u$-derivative of
the integral (\ref{bmass}), using suitable vacuum
Einstein's equations. However, since we wish to account
for the mass-loss in terms of the physical degrees of freedom
$\rho_{a b}$, we shall work with the volume integral (\ref{ha1}).

Recalling that diff$N_{2}$ is the residual gauge symmetry of the
metric (\ref{res}), we may fix this symmetry
by choosing $A_{+}^{\ a}=0$\footnote{This is
equivalent to the assumption that $\partial/ \! \partial u$
is a {\it twist-free} null vector field\cite{yoon,yoona,exact}.
However, there could be some topological obstructions against
globalizing this choice.}
by a suitable coordinate transformation
on $N_{2}$. Then the equation $C_{+}=0$ in (\ref{geq})
reduces to the following Schr\"{o}dinger equation
\begin{equation}
\partial_{+}^{2}{\rm e}^{ \sigma / 2}+ \kappa_{+}^{' 2}
{\rm e}^{ \sigma / 2}=0,  \hspace{.5cm}
{\rm where}\hspace{.5cm}
\kappa_{+}^{' 2}:={1\over 8}\rho^{a b}\rho^{c d}
 (\partial_{+}\rho_{a c})
 (\partial_{+}\rho_{b d})\geq 0.                \label{geqa}
\end{equation}
In the following we shall use this equation when we
examine the rate of change in $E$ as
the retarded time $u$ advances.
Let us further notice that the metric
(\ref{res}) becomes, in the gauge $A_{+}^{\ a}=0$,
\begin{equation}
ds^{2}=-2du dv + {\rm e}^{ \sigma }
\rho_{a b}(A_{-}^{\ a}dv + dy^{a})
(A_{-}^{\ b}dv + dy^{b}).                    \label{lc}
\end{equation}
With $y^{a}=(\vartheta, \varphi)$, where
$\vartheta, \varphi$ are the angles of $S_{2}$,
we find by comparing (\ref{lc}) with (\ref{null})
the following asymptotic behaviors of the metric
as $v$ approaches to infinity,
\begin{equation}
{\rm e}^{ \sigma }=O(v^{2}), \hspace{.5cm}
\rho_{a b}=O(1), \hspace{.5cm}
A_{-}^{\ a}=O(1/ v^{2}).               \label{asym}
\end{equation}
In the gauge $A_{+}^{\ a}=0$\footnote{This gauge
choice is only for convenience, since we already obtained the
covariant expression of the Bondi mass in the previous
subsection.},
the volume integral (\ref{ha1}) becomes
\begin{eqnarray}
E&=&\int dv  d^{2} y \ {\rm e}^{\sigma} \Big\{
  R_{2}+ 2(\partial_{+}\sigma)(D_{-}\sigma)
   + 2 \partial_{+}D_{-}\sigma  \Big\}    \nonumber\\
 &=&\int dv  d^{2}y \Big\{ {\rm e}^{\sigma} R_{2}
  + 2\partial_{+}D_{-}{\rm e}^{\sigma} \Big\},  \label{ha10}
\end{eqnarray}
where we used the identity
\begin{equation}
\partial_{+}D_{-}{\rm e}^{\sigma}
= {\rm e}^{\sigma}(\partial_{+}\sigma)(D_{-}\sigma)
+{\rm e}^{\sigma}\partial_{+}D_{-}\sigma.             \label{sug}
\end{equation}
The first term in (\ref{ha10}) is the $v$-integration of the
Euler number $\chi$, where
\begin{equation}
\chi ={1\over 4\pi}\int d^{2}y \ {\rm e}^{\sigma}
R_{2}=2, 0, -2(g-1),
\end{equation}
for $N_{2}=S_{2}$, $T_{2}$, and the 2-surface
$\Sigma_{g}$ of genus $g$, respectively.
Since the $u$-derivative of the Euler integral
is zero, the rate of change in $E$ as $u$ increases comes from the
second term in (\ref{ha10}). For the asymptotically flat
space-times, we may assume $N_{2}=S_{2}$ so that
\begin{eqnarray}
{d E\over  d u}&=&
   2\int_{S_{2}} dv d^{2}y \
   \partial_{+}^{2} (D_{-}{\rm e}^{\sigma})        \nonumber\\
&=&2\int_{v=\infty, S_{2}} \!\!\!\!\!\! d^{2}y \
   (\partial_{+}^{2}{\rm e}^{\sigma})
  -2\int_{v=v_{0}, S_{2}} \!\!\!\!\!\! d^{2}y \
   (\partial_{+}^{2} {\rm e}^{\sigma})
   \Big(  1 + O(v_{0}^{-1}) \Big),
\end{eqnarray}
where the domain of the $v$-integration was chosen from
$v_{0}$ to $\infty$, and we used the asymptotic behaviors
(\ref{asym}). Here $v_{0}$ is some point that
lies sufficiently far away from the sources of
gravitational waves along the out-going null
direction such that the gravitational waves are contained entirely
in the range $v_{0}< v \leq \infty$ at the instant
$u={\rm constant}$. In this asymptotic region,
we may also assume the
out-going null condition\cite{bondi,sach,newmantod},
\begin{equation}
{\rm e}^{\sigma}={1\over 2}v^{2}{\rm sin}\vartheta
\Big\{ 1 + {f(u,\vartheta,\varphi) \over  v}
+ O( {1\over v^{2}} ) \Big\},          \label{out}
\end{equation}
for some function $f(u,\vartheta,\varphi)$.
{}From this, it is found that
\begin{eqnarray}
\partial_{+}^{2}{\rm e}^{\sigma}&=&2{\rm e}^{\sigma / 2}
\partial_{+}^{2}{\rm e}^{\sigma /  2} \
\Big( 1+ O(1/  v) \Big)                             \nonumber\\
&=&-{1\over 4}{\rm e}^{\sigma}\rho^{a b}\rho^{c d}
   (\partial_{+}\rho_{a c})
   (\partial_{+}\rho_{b d})\Big( 1+ O(1/  v) \Big),   \label{as}
\end{eqnarray}
where in the second line we used the Schr\"{o}dinger equation
(\ref{geqa}). Thus $d E/ \! du$ becomes
\begin{eqnarray}
{d E\over  d u}&=&-{1\over 2}
   \lim_{v\rightarrow\infty}\int_{S_{2}} d^{2}y \
   {\rm e}^{\sigma}\rho^{a b}\rho^{c d}
   (\partial_{+}\rho_{a c})(\partial_{+}\rho_{b d}) \nonumber\\
& &  +{1\over 2}\int_{v=v_{0}, S_{2} }\!\!\!\!\!\! d^{2}y \
  {\rm e}^{\sigma}\rho^{a b}\rho^{c d}(\partial_{+}\rho_{a c})
  (\partial_{+}\rho_{b d})
  \Big( 1+ O(v_{0}^{-1})\Big).             \label{ala}
\end{eqnarray}
Since there are no
propagating gravitational degrees of freedom
in the region $v\leq v_{0}$ the {\it currents} of
gravitational waves must vanish in this region, so that
\begin{equation}
\rho^{a b}\partial_{+}\rho_{a c}=0 \hspace{.5cm}
{\rm for} \ \ \ v\leq v_{0}.        \label{curr}
\end{equation}
Thus the second term in (\ref{ala}) vanishes, and
we finally have
\begin{equation}
{d E \over  d u}
=-{1\over 2}\lim_{v\rightarrow\infty}\int_{ S_{2} } d^{2}y \
{\rm e}^{\sigma}\rho^{a b}\rho^{c d}
(\partial_{+}\rho_{a c})
(\partial_{+}\rho_{b d}) \leq 0.                \label{bondi}
\end{equation}
Apart from the integration measure ${\rm e}^{\sigma}$, this
flux integral over $S^{2}$ at null infinity is expressed
entirely in terms of the physical degrees of freedom, and
is negative-definite. This is precisely the Bondi
mass-loss formula!
It must be stressed that the gravitational energy
carried away to null infinity by the gravitational radiation is
given in a bilinear combination of the gravitational currents,
in excellent accordance with our experience that observables
are very often expressed in bilinears of the physical fields.
This strongly supports the view held by the geometric
quantization school\cite{wood,ish,git} that, in general relativity,
as in other non-linear field
theories, it is the current rather than the conformal 2-geometry
that should be regarded as the fundamental physical field.

The total radiated gravitational energy to null infinity
between the null time interval $u_{0}$ and $u$ can be obtained
by integrating (\ref{bondi}), and is given by
\begin{equation}
E(u) - E(u_{0})=-{1\over 2}\lim_{v\rightarrow\infty}
\int_{u_{0}}^{u} du \int_{S^{2}} d^{2}y \
{\rm e}^{\sigma}\rho^{a b}\rho^{c d}
(\partial_{+}\rho_{a c})(\partial_{+}\rho_{b d}). \label{bondib}
\end{equation}

\subsection{The ADM Mass}
\label{subsec:adm}

Now we shall show that the volume integral of
the proposed local gravitational energy density
(\ref{totalham}) over a spacelike hypersurface
reproduces the ADM surface integral.
Let us make the following coordinate transformation
\begin{equation}
r=-{1\over 2}u + v.\label{r}
\end{equation}
In the new coordinates $(u,r, y^{a})$, the metric
(\ref{res}) becomes
\begin{eqnarray}
ds^{2} &=&- du^{2} -2dudr  + {\rm e}^{\sigma}\rho_{a b} \Big\{
  (A_{+}^{\ a} + {1\over 2}A_{-}^{\ a}) du
  + A_{-}^{\ a} dr + dy^{a}\Big\}                  \nonumber\\
& &  \Big\{ (A_{+}^{\ b} + {1\over 2}A_{-}^{\ b}) du
  + A_{-}^{\ b} dr + dy^{b}\Big\}.         \label{ra}
\end{eqnarray}
Since we still have the residual gauge symmetry
associated with the
diff$N_{2}$ invariance in (\ref{ra}), we may well fix this
residual symmetry by choosing
\begin{equation}
A_{+}^{\ a} + {1\over 2}A_{-}^{\ a}=0.
\end{equation}
Then (\ref{ra}) reduces to
\begin{equation}
ds^{2} =- du^{2} -2dudr  + {\rm e}^{\sigma}\rho_{a b}
   \Big( A_{-}^{\ a} dr + dy^{a}\Big)
   \Big( A_{-}^{\ b} dr + dy^{b}\Big).         \label{ral}
\end{equation}
In the limit as $r\rightarrow \infty$, both (\ref{ra})
and (\ref{ral}) approach over to the flat space-time metric
\begin{equation}
ds^{2} \rightarrow - du^{2} -2dudr  + {1\over 2} r^{2}
(d\vartheta ^{2} +
{\rm sin}^{2}\vartheta d\varphi^{2}),                 \label{raa}
\end{equation}
showing that $r/ \sqrt{2}$ becomes the area radius
in this limit. Moreover the $u$-coordinate is the proper time
at each point on the $u={\rm constant}$
hypersurface,
suggesting that the volume integral be defined over the
spacelike hypersurface $u={\rm constant}$ in the new coordinates,
since the ADM surface integral is associated with
a unit time translation at spatial infinity.

To find the relevant Hamiltonian for the ADM mass we need
to write the local Lagrangian density in the
new coordinates. The local Lagrangian density can be
found directly from (\ref{finala})
\begin{equation}
{\cal L}={1\over 2}{\rm e}^{2\sigma}\rho_{a b}
  F_{+-}^{\ \ a}F_{+-}^{\ \ b}
 -{\rm e}^{\sigma}(D_{+}\sigma) (D_{-}\sigma)
 +{1\over 2}{\rm e}^{\sigma}\rho^{a b}\rho^{c d}
 (D_{+}\rho_{a c})(D_{-}\rho_{b d})
 + \sum_{\alpha = \pm, 0}
 \lambda^{\alpha}C_{\alpha}
  \mid_{\partial_{-}=\partial_{r}},              \label{hec}
\end{equation}
with $\partial_{-}$ replaced by $\partial_{r}$ everywhere.
Thus the relevant volume integral is given by
\begin{eqnarray}
E&=&{1\over 2}\int dr d^{2} y \ {\rm e}^{2\sigma}\rho_{a b}
    F_{+-}^{\ \ a}F_{+-}^{\ \ b}
    \mid_{\partial_{-}=\partial_{r}}               \nonumber\\
&=&{1\over 3}\int dr  d^{2} y \ {\rm e}^{\sigma} \Big\{
   R_{2}+ (D_{+}\sigma)
   (D_{-}\sigma) - {1\over 2}\rho^{a b}\rho^{c d}
   (D_{+}\rho_{a c})(D_{-}\rho_{b d})
   \Big\}\mid_{\partial_{-}=\partial_{r}}          \nonumber\\
&=&{1\over 3}\int dr d^{2} y \ {\rm e}^{\sigma }
   h^{A C}h^{B D} C_{A B C D}.                   \label{newha1}
\end{eqnarray}
Repeating the same reasoning as in Appendix C,
with $\tilde{\Omega}^{-1}=r$, we find that the volume integral
(\ref{newha1}) becomes the surface integral,
\begin{equation}
E={1\over 3} \lim \ r\!\! \int d^{2}y \ {\rm e}^{\sigma}
h^{A C}h^{B D} C_{A B C D},            \label{adma}
\end{equation}
which becomes precisely the covariant expression of
the ADM mass of the asymptotically flat space-times in the limit
as $r$ approaches to infinity!\cite{haya,ash,asha,ashb}

\section{Discussions}
\label{sec:dis}

In this article, we combined the double null formalism of
general relativity with the KK formalism in the (2,2)-splitting,
and proposed a local gravitational energy
density of general relativity. As we have seen so far,
there are a number of notable features of this description
which deserve further remarks.
First of all, this formalism explicitly brings out the
gauge theory aspects of general relativity of the 4-dimensional
space-times. Although it has been realized for
a long time that the local diffeomorphism invariance in general
relativity is on a par with the local gauge symmetry
in gauge theories, it seems fair to say that
the full-fledged gauge theory
formulation of general relativity is still lacking.
Our 4-dimensional KK approach to general
relativity in the (2,2)-splitting, using the double null
gauge, seems to provide such a formulation, as we have
described in this article. Thus we may well take care of the
Gauss-law equations associated with the diff$N_{2}$ invariance
by considering the diff$N_{2}$ invariant quantities only.

Moreover, this formalism shows that, in the double null gauge,
local gravitational energy
density of general relativity can be well-defined, and moreover,
is positive-definite. The volume integral of this
local gravitational energy density over the
3-dimensional null and spacelike hypersurface correctly
reproduces the Bondi and ADM surface integral at null and spatial
infinity, respectively.
The Bondi mass-loss due to the gravitational radiation
in the asymptotically flat space-times is given by a
negative-definite flux integral of
the bilinear in the gravitational currents
at null infinity. It should be mentioned that
the proposed gravitational energy density can
be also used to define quasi-local
gravitational energies  for a finite region of a given 3-dimensional
hypersurface in a straightforward
way\cite{haya,haw,hor,sch,berg,david}.

The non-zero local Hamiltonian density proposed in this article
also has a direct bearing to the problem of
time\cite{dirac,jim,karel,car,rov}. The time associated
with the non-zero Hamiltonian is the retarded time
$u$, which has the physical meaning as the phase of the
gravitational radiation when gravitational waves are present.
The canonical analysis of our formalism is under
progress\cite{taejin}, which will shed further light on
this important issue.

In addition, this
formalism does seem to indicate the intriguing possibility
that quantum general relativity of the 4-dimensional
space-time may be regarded as a (1+1)-dimensional
quantum field theory. For instance, one might even speculate
that quantum gravity might be a finite theory,
given that the renormalizability depends critically
on the dimensions of ``space-time". However, it must be addressed
that this formalism is for the vacuum general relativity only.
It certainly is an interesting question to see whether this
formalism can be extended to include matter fields.
We leave this problem for the future investigation.

\acknowledgments

It is a great pleasure to thank G.T. Horowitz for the hospitality
during the Space-Time 93 program at ITP, Santa Barbara,
where part of this work was conceived. The author also
thanks S. Carlip,
C.W. Misner, V. Moncrief, E.T. Newman and others for their
interest in this approach and for encouragements, and
K. Kucha\v{r} for suggesting him to look into the null hypersurface
formalism that was of invaluable help to this work. He also
thanks D. Brill for a number of valuable suggestions on the
preliminary version of this article, and for kindly informing
him the related work of S.A. Hayward, and for encouragements.
This work is supported in part by the
Ministry of Education and by the Korea Science and Engineering
Foundation through the SRC program.

\appendix
\section{The E-H Lagrangian Density in the General (2,2)-Splitting}

In this appendix, we shall make a general (2,2)-decomposition of
space-time, and obtain the corresponding E-H Lagrangian
density\cite{soh,yoon,yoona}
without picking up a particular gauge. By examining the
transformation properties of the metric in the (2,2)-splitting
under the diff$N_{2}$ transformations, we shall find
that each field can be identified either as a scalar, a tensor,
or a gauge field with respect to the diff$N_{2}$ transformations,
respectively, suggesting that the KK program
works even in the absence of any Killing vector fields.
Then we simplify the general E-H Lagrangian density
by introducing the double null gauge to obtain ${\cal L}_{0}$ in
(\ref{ga})\cite{yoona}, which also has the
diff$N_{2}$ symmetry as the residual symmetry.

The 4-dimensional space-time may be regarded as a fibred manifold,
i.e. a local product of two 2-dimensional submanifolds
$M_{1+1} \times N_{2}$, for which we introduce two pairs of
the basis vector fields
$\partial_{\mu}=\partial / \!  \partial x^{\mu}$$ (\mu=0,1)$ and
$\partial_{a}=\partial / \!  \partial y^{a}$$ (a=2,3)$,
respectively. The corresponding metrics on $M_{1+1}$ and
$N_{2}$ will be denoted as $\gamma_{\mu\nu}$ and $\phi_{a b}$,
respectively. Then the general
line element of the 4-dimensional space-time can be written as
\renewcommand{\theequation}{A\arabic{equation}}
\setcounter{equation}{0}
\begin{equation}
ds^{2}=\gamma_{\mu\nu}dx^{\mu}dx^{\nu}
+ \phi_{a b}(A_{\mu}^{\ a}dx^{\mu} + dy^{a})
            (A_{\nu}^{\ b}dx^{\nu} + dy^{b}).     \label{gen}
\end{equation}
Formally this is quite similar to the ``dimensional reduction" in
KK theory, where $M_{1+1}$ is regarded as the
(1+1)-dimensional ``space-time" and $N_{2}$ as
the ``internal" fibre space. In the standard KK reduction
one assumes
a restriction on the metric, namely, an isometry condition,
to make $A_{\mu}^{\ a}$ a gauge field associated
with the isometry group. Here, however, we do not assume
such isometry conditions,
and allow all the fields to depend on both $x^\mu $
and $y^a $.
Nevertheless  $A_{\mu}^{\ a}(x,y)$ can still be
identified as a gauge field,
but now associated with an infinite dimensional diffeomorphism
group diff$N_{2}$. To show this, let us consider the
following diffeomorphism of $N_{2}$,
\begin{equation}
y^{' a}=y^{' a} (x,y),
\hspace{1cm}
x^{' \mu}=x^{\mu}.                      \label{new}
\end{equation}
Under this transformation, we find
\begin{eqnarray}
& &\gamma '_{\mu\nu}(x,y')=\gamma _{\mu\nu}(x,y),   \label{gam}\\
& &\phi'_{a b}(x,y')={\partial y^{c} \over \partial y^{' a}}
   {\partial y^{d} \over \partial y^{' b}}
   \phi_{c d}(x,y),                          \label{phi}\\
& &A_{\mu}^{\ 'a}(x,y')={\partial y^{' a} \over \partial y^{c}}
   A_{\mu}^{\ c}(x,y) -\partial_{\mu} y^{' a}, \label{newtran}
\end{eqnarray}
such that the line element (\ref{gen}) is invariant.
Under the corresponding infinitesimal transformation
\begin{equation}
\delta y^{a}=\xi^{a} (x,y),   \label{var}
\hspace{1cm}
\delta x^{\mu}=0,
\end{equation}
these become
\begin{eqnarray}
\delta \gamma_{\mu\nu}&=&-[ \xi, \gamma_{\mu\nu}]
=-\xi^{c}\partial_{c} \gamma_{\mu\nu},      \label{var3}\\
\delta \phi_{a b}&=&-[ \xi, \phi ]_{a b}
=-\xi^{c}\partial_{c}\phi_{a b}
   -(\partial_{a}\xi^{c})\phi_{c b}
   -(\partial_{b}\xi^{c})\phi_{a c},   \label{var2}\\
\delta A_{\mu}^{\ a}&=&-\partial_{\mu}\xi^{a} + [A_{\mu},\ \xi ]^a
=-\partial_{\mu}\xi^{a}+(A_{\mu}^{c}
    \partial_{c}\xi^{a}
    -\xi^{c}\partial_{c}A_{\mu}^{\ a}), \label{var1}
\end{eqnarray}
where the bracket represents the Lie derivative that acts on the
``internal" indices $a, b,$ etc, only.  Notice that the
Lie derivative, an {\it infinite}
dimensional generalization of the finite dimensional matrix
commutators, appears naturally.
Associated with this diff$N_{2}$ transformation, the
diff$N_{2}$-covariant derivative $D_{\mu}$ is defined by
\begin{equation}
D_{\mu}=\partial_{\mu}-[A_{\mu}, \ \ ],       \label{cd}
\end{equation}
where the bracket is again the Lie derivative along
$A_{\mu}=A_{\mu}^{\ a}\partial_{a}$. With this definition,
we have
\begin{equation}
\delta A_{\mu}^{\ a}=-D_{\mu}\xi^{a},
\end{equation}
which clearly indicates that  $A_{\mu}^{\ a}$ is the
gauge field valued in
the infinite dimensional Lie algebra associated with
diff$N_{2}$. Moreover
the transformation properties
(\ref{var3}) and (\ref{var2}) show that $\gamma_{\mu\nu}$ and
$\phi_{a b}$ are a scalar and a tensor field, respectively, under
diff$N_{2}$.
The field strength $F_{\mu\nu}^{\ \ a}$ corresponding
to $A_{\mu}^{\ a}$ can now be defined as
\begin{equation}
[D_\mu, D_\nu]=-F_{\mu\nu}^{\ \ a}\partial_a=-\{
  \partial_\mu A_{\nu} ^ { \ a}-\partial_\nu
A_{\mu} ^ { \ a} - [A_{\mu}, A_{\nu}]^a \} \partial_a, \label{fiea}
\end{equation}
which transforms covariantly under the
infinitesimal transformation (\ref{var}),
\begin{equation}
\delta F_{\mu\nu}^{\ \ a}=-[\xi, F_{\mu\nu}]^{a}.
\end{equation}
To obtain the E-H Lagrangian density, we have to first
compute connections and curvature tensors.
For this purpose it is convenient to
introduce the following horizontal lift basis
\begin{math}
\hat{\partial}_{A} = (\hat{\partial}_{\mu} , \hat{\partial}_{a} )
\end{math}
where\cite{cho}
\begin{equation}
\hat{\partial}_\mu := \partial_\mu
- A_{\mu} ^ {\ a}\partial_a,
\hspace{1cm}
\hat{\partial}_a :=\partial_a  \ .  \label{cova}
\end{equation}
{}From the following commutation relations
\begin{equation}
[\hat{\partial}_A, \hat{\partial}_B]=f_{A B} ^ { \
\  \ C}\hat{\partial}_C,
\end{equation}
we find the structure {\it functions} $f_{A B} ^ { \  \  \ C}(x,y)$
\begin{eqnarray}
& &f_{\mu\nu} ^ { \  \ a}
   =-F_{\mu\nu} ^ { \  \ a},                  \nonumber\\
& &f_{\mu a} ^ { \  \ b}=-f_{a \mu} ^ { \  \ b}
  =\partial_a A_{\mu} ^ { \ b},                 \nonumber\\
& &f_{A B}^{ \ \ \ C}=0, \hspace{1cm} {\rm otherwise}.\label{str}
\end{eqnarray}
The virtue of this basis is that it brings the metric (\ref{gen})
into a block diagonal form
\begin{equation}
\hat{g}_{A B}=\left(\matrix{\gamma_{\mu\nu}  &
0 \cr 0 & \phi_{ab} \cr }\right),
\end{equation}
which drastically simplifies the computation of the
scalar curvature.

The connection coefficients and the curvature tensors in
this basis are given by\cite{cho,mtw}
\begin{eqnarray}
& &\hat{\Gamma}_{A B}^{ \ \ \ C}={1\over 2}\hat{g}^{C D}
  \Big( \hat{\partial}_{A}\hat{g}_{B D}
  +\hat{\partial}_{B}\hat{g}_{A D}
  - \hat{\partial}_{D}\hat{g}_{A B} \Big)
  +{1\over 2}\hat{g}^{C D}\Big( f_{A B D} - f_{B D A}
 - f_{A D B} \Big),                                \nonumber\\
& &\hat{R}_{A B C}^{ \ \ \ \ \ D}
  = \hat{\partial}_{B}\hat{\Gamma}_{A C}^{\ \ \ D}
  -\hat{\partial}_{A}\hat{\Gamma}_{B C}^{\ \ \ D}
  + \hat{\Gamma}_{B E}^{\ \ \ D}\hat{\Gamma}_{A C}^{\ \ \ E}
  - \hat{\Gamma}_{A E}^{\ \ \ D}\hat{\Gamma}_{B C}^{\ \ \ E}
  +f_{A B}^{\ \ \ E}\hat{\Gamma}_{E C}^{\ \ \ D},  \nonumber\\
& &\hat{R}_{A C}=\hat{g}^{B D}\hat{R}_{A B C D}, \nonumber\\
& &R=\hat{g}^{A C}\hat{R}_{A C},
\end{eqnarray}
where $f_{A B C}:=\hat{g}_{C D}f_{A B}^{\ \ \ D}$.
In components the connection coefficients are given by
\begin{eqnarray}
& &\hat{\Gamma}_{\mu\nu}^{\ \ \alpha}
   ={1\over 2}\gamma^{\alpha\beta}\Big(
   \hat{\partial}_{\mu}\gamma_{\nu\beta}
   + \hat{\partial}_{\nu}\gamma_{\mu\beta}
   -\hat{\partial}_{\beta}\gamma_{\mu\nu}  \Big), \nonumber\\
& &\hat{\Gamma}_{\mu\nu}^{\ \ a}
   = -{1\over 2}\phi^{a b}\partial_{b}\gamma_{\mu\nu}
  - {1\over 2}F_{\mu\nu}^{\ \ a},                \nonumber\\
& &\hat{\Gamma}_{\mu a}^{\ \ \nu}=\hat{\Gamma}_{ a\mu}^{\ \ \nu}
   ={1\over 2}\gamma^{\nu\alpha}\partial_{a}\gamma_{\mu\alpha}
   + {1\over 2}\gamma^{\nu\alpha}
    \phi_{a b}F_{\mu\alpha}^{\ \ b}, \         \nonumber\\
& &\hat{\Gamma}_{\mu a}^{\ \ b}
   ={1\over 2}\phi^{b c}\hat{\partial}_{\mu}\phi_{a c}
   +{1\over 2}\partial_{a}A_{\mu}^{\ b}
   - {1\over 2}\phi^{b c}
    \phi_{a e}\partial_{c}A_{\mu}^{\ e},           \nonumber\\
& &\hat{\Gamma}_{a\mu}^{\ \ b}
   ={1\over 2}\phi^{b c}\hat{\partial}_{\mu}\phi_{a c}
   -{1\over 2}\partial_{a}A_{\mu}^{\ b}
   -{1\over 2}\phi^{b c}\phi_{a e}
   \partial_{c}A_{\mu}^{\ e},                     \nonumber\\
& &\hat{\Gamma}_{a b}^{\ \ \mu}
  =-{1\over 2}\gamma^{\mu\nu}\hat{\partial}_{\nu}\phi_{a b}
  +{1\over 2}\gamma^{\mu\nu}\phi_{a c}\partial_{b}A_{\nu}^{\ c}
  +{1\over 2}\gamma^{\mu\nu}\phi_{b c}
   \partial_{a}A_{\nu}^{\ c},                      \nonumber\\
& &\hat{\Gamma}_{a b}^{\ \ c}={1\over 2}\phi^{c d}\Big(
  \partial_{a}\phi_{b d} + \partial_{b}\phi_{a d}
  - \partial_{d}\phi_{a b}\Big),
\end{eqnarray}
The following identities are also useful;
\begin{eqnarray}
& &\hat{\Gamma}_{\mu\nu}^{\ \ \mu}
  ={1\over 2}\gamma^{\alpha\beta}
   \hat{\partial}_{\nu}\gamma_{\alpha\beta}, \hspace{1cm}
\hat{\Gamma}_{\beta a}^{\ \ \beta}
  ={1\over 2}\gamma^{\alpha\beta}
  \partial_{a}\gamma_{\alpha\beta},          \hspace{1cm}
\hat{\Gamma}_{a \nu}^{\ \ a}={1\over 2}\phi^{a b}
 \hat{\partial}_{\nu}\phi_{a b}
  -\partial_{a}A_{\nu}^{\ a},        \nonumber\\
& &\hat{\Gamma}_{\nu a}^{\ \ a}
  ={1\over 2}\phi^{a b}\hat{\partial}_{\nu}\phi_{a b},
\hspace{1cm}
\hat{\Gamma}_{a b}^{\ \ a}
={1\over 2}\phi^{a c}\partial_{b}\phi_{a c}. \label{use}
\end{eqnarray}
The Ricci tensors and the scalar curvature are given by
\begin{equation}
\hat{R}_{\mu\nu}=\hat{R}_{\mu \alpha \nu}^{\ \ \ \ \alpha}
+\hat{R}_{\mu a \nu}^{\ \ \ \ a}, \hspace{1cm}
\hat{R}_{a c}=\hat{R}_{a b c}^{\ \ \ \ b}
+\hat{R}_{a \alpha c}^{\ \ \ \ \alpha}, \hspace{1cm}
R=\gamma^{\mu\nu}\hat{R}_{\mu\nu}
 + \phi^{a c}\hat{R}_{a c}.              \label{lll}
\end{equation}
Thus, in order to obtain the E-H Lagrangian density, we
need to calculate $\gamma^{\mu\nu}\hat{R}_{\mu\nu}$ and
$\phi^{a c}\hat{R}_{a c}$ only. Let us first define
$R_{\mu\nu}'$ and $R_{a c}$ as follows;
\begin{eqnarray}
& &R_{\mu\nu}'= \hat{\partial}_{\alpha}
  \hat{\Gamma}_{\mu \nu}^{\ \ \alpha}
 -\hat{\partial}_{\mu}\hat{\Gamma}_{\alpha \nu}^{\ \ \alpha}
 +\hat{\Gamma}_{\beta \alpha}^{\ \ \beta}
  \hat{\Gamma}_{\mu \nu}^{\ \ \alpha}
 -\hat{\Gamma}_{\mu \beta}^{\ \ \alpha}
  \hat{\Gamma}_{\alpha \nu}^{\ \ \beta},          \nonumber\\
& &R_{a c}=\partial_{b}\hat{\Gamma}_{a c}^{\ \ b}
-\partial_{a}\hat{\Gamma}_{b c}^{\ \ b}
+\hat{\Gamma}_{d b}^{\ \ d}\hat{\Gamma}_{a c}^{\ \ b}
-\hat{\Gamma}_{a d}^{\ \ b}
\hat{\Gamma}_{b c}^{\ \ d}.                 \label{qten}
\end{eqnarray}
Notice that formally $R_{\mu\nu}'$ is identical to the
Ricci tensor of $M_{1+1}$, except that
$\hat{\partial}_{\mu}$ was used
instead of $\partial_{\mu}$.  For this reason it
might be called  the ``gauged" Ricci tensor of $M_{1+1}$,
whereas $R_{a c}$ is the usual Ricci tensor of $N_{2}$.
After a long computation we obtain
$\gamma^{\mu\nu}\hat{R}_{\mu\nu}$
and  $\phi^{a c}\hat{R}_{a c}$ as follows;
\begin{eqnarray}
\gamma^{\mu\nu}\hat{R}_{\mu\nu}&=&\gamma^{\mu\nu}R_{\mu\nu}'
-{1\over 2}\gamma^{\mu\nu}\gamma^{\alpha\beta}\phi_{a b}
  F_{\mu\alpha} ^ { \  \ a} F_{\nu\beta}^{\ \ b}
-{1\over 4}\gamma^{\mu\nu}\phi ^ {a b}\phi ^ {c d}
 (\hat{\partial}_{\mu}\phi_{a c})
 (\hat{\partial}_{\nu}\phi_{b d})          \nonumber\\
& &+{1\over 4}\gamma^{\mu\nu}\phi ^ {a b}\phi ^ {c d}
  (\hat{\partial}_{\mu}\phi_{a b})
  (\hat{\partial}_{\nu}\phi_{c d})
+\gamma^{\mu\nu}\phi^{b c}(\hat{\partial}_{\mu}\phi_{a c})
 (\partial_{b}A_{\nu}^{\ a})                  \nonumber\\
& &-\gamma^{\mu\nu}\phi^{a b}(\hat{\partial}_{\mu}\phi_{a b})
 (\partial_{c}A_{\nu}^{\ c})
  +\gamma^{\mu\nu}(\partial_{a}A_{\mu}^{\ a})
 (\partial_{b}A_{\nu}^{\ b})
-{1\over 2}\gamma^{\mu\nu}(\partial_{a}A_{\mu}^{\ b})
 (\partial_{b}A_{\nu}^{\ a})                    \nonumber\\
& &-{1\over 2}\gamma^{\mu\nu}\phi ^ {a b}\phi _ {c d}
 (\partial_{a}A_{\mu}^{\ c})
 (\partial_{b}A_{\nu}^{\ d})        \nonumber\\
& &-(\hat{\nabla}_{\mu}+\hat{\Gamma}_{c \mu}^{\ \ c})
 \Big( {1\over 2}\gamma^{\mu\nu}\phi^{a b}
 \hat{\partial}_{\nu}\phi_{a b}
  -\gamma^{\mu\nu}\partial_{a}A_{\nu}^{\ a}\Big)  \nonumber\\
& &-(\hat{\nabla}_{a}+\hat{\Gamma}_{\alpha a}^{\ \ \alpha})
 \Big( {1\over 2}\phi^{a b} \gamma^{\mu\nu}
 \partial_{b}\gamma_{\mu\nu}\Big),                \label{qqq}\\
\phi^{a c}\hat{R}_{a c}&=&\phi^{a c}R_{a c}
+{1\over 4}\gamma^{\mu\nu}\gamma^{\alpha\beta}
  \phi_{a b}F_{\mu\alpha} ^ { \  \ a}
  F_{\nu\beta}^{\ \ b}
-{1\over 4}\phi ^ {a b}\gamma^{\mu\nu}\gamma^{\alpha\beta}
        (\partial_{a}\gamma_{\mu \alpha})
 (\partial_{b}\gamma_{\nu\beta})               \nonumber\\
& &+{1\over 4}\phi ^ {a b}\gamma^{\mu\nu}\gamma^{\alpha\beta}
(\partial_{a}\gamma_{\mu\nu})
(\partial_{b}\gamma_{\alpha\beta})           \nonumber\\
& &-(\hat{\nabla}_{\mu}+\hat{\Gamma}_{c \mu}^{\ \ c})
 \Big( {1\over 2}\gamma^{\mu\nu}\phi^{a b}
  \hat{\partial}_{\nu}\phi_{a b}
- \gamma^{\mu\nu}\partial_{a}A_{\nu}^{\ a}\Big)  \nonumber\\
& &-(\hat{\nabla}_{a}+\hat{\Gamma}_{\alpha a}^{\ \ \alpha})
 \Big( {1\over 2}\phi^{a b} \gamma^{\mu\nu}
 \partial_{b}\gamma_{\mu\nu}\Big).               \label{rrr}
\end{eqnarray}
Here the derivatives $\hat{\nabla}_{\mu}$ and
$\hat{\nabla}_{a}$ are compatible with the metrics
$\gamma_{\alpha\beta}$ and $\phi_{b c}$ in the
horizontal lift basis, respectively, such that
\begin{eqnarray}
& &\hat{\nabla}_{\mu}\gamma_{\alpha\beta}
=\hat{\partial}_{\mu}\gamma_{\alpha\beta}
-\hat{\Gamma}_{\mu\alpha}^{\ \ \delta}\gamma_{\delta\beta}
-\hat{\Gamma}_{\mu\beta}^{\ \ \delta}
\gamma_{\alpha\delta}=0,     \nonumber\\
& &\hat{\nabla}_{a}\phi_{b c}=\partial_{a}\phi_{b c}
  -\hat{\Gamma}_{a b}^{\ \ d}\phi_{d c}
  -\hat{\Gamma}_{a c}^{\ \ d}\phi_{b d}=0.
\end{eqnarray}
For an object of the mixed-type such as
$X_{a\cdots\alpha\cdots}^{\hspace{.2cm}b\cdots\beta\cdots}$
in this basis, these derivatives act as follows;
\begin{eqnarray}
& &\hat{\nabla}_{\mu}
X_{b\cdots\alpha\cdots}^{\hspace{.2cm}c\cdots\beta\cdots}
=\hat{\partial}_{\mu}
X_{b\cdots\alpha\cdots}^{\hspace{.2cm}c\cdots\beta\cdots}
-\hat{\Gamma}_{\mu\alpha}^{\ \ \delta}
X_{b\cdots\delta\cdots}^{\hspace{.2cm}c\cdots\beta\cdots}
+\hat{\Gamma}_{\mu\delta}^{\ \ \beta}
  X_{b\cdots\alpha\cdots}^{\hspace{.2cm}c\cdots\delta\cdots}
+\cdots ,                   \nonumber\\
& &\hat{\nabla}_{a}
X_{b\cdots\alpha\cdots}^{\hspace{.2cm}c\cdots\beta\cdots}
=\partial_{a}
X_{b\cdots\alpha\cdots}^{\hspace{.2cm}c\cdots\beta\cdots}
-\hat{\Gamma}_{a b}^{\ \ d}
X_{d\cdots\alpha\cdots}^{\hspace{.2cm}c\cdots\beta\cdots}
+\hat{\Gamma}_{a d}^{\ \ c}
X_{b\cdots\alpha\cdots}^{\hspace{.2cm}d\cdots\beta\cdots}
+ \cdots.                      \label{newd}
\end{eqnarray}
For instance, the followings are true,
\begin{eqnarray}
& &\hat{\nabla}_{\mu}(\phi^{a b}\hat{\partial}_{\nu}\phi_{a b})
=\hat{\partial}_{\mu}(\phi^{a b}\hat{\partial}_{\nu}\phi_{a b})
-\hat{\Gamma}_{\mu\nu}^{\ \ \alpha}(\phi^{a b}
\hat{\partial}_{\alpha}\phi_{a b}),             \nonumber\\
& &\hat{\nabla}_{a}(\gamma^{\mu\nu}\partial_{b}\gamma_{\mu\nu})
=\partial_{a}(\gamma^{\mu\nu}\partial_{b}\gamma_{\mu\nu})
-\hat{\Gamma}_{a b}^{\ \ c}(\gamma^{\mu\nu}
\partial_{c}\gamma_{\mu\nu}), \nonumber\\
& &\hat{\nabla}_{\mu}(\partial_{a}A_{\nu}^{\ a})
 =\hat{\partial}_{\mu}(\partial_{a}A_{\nu}^{\ a})
 -\hat{\Gamma}_{\mu\nu}^{\ \ \alpha}
  \partial_{a}A_{\alpha}^{\ a},
\end{eqnarray}
which we used in (\ref{qqq}) and (\ref{rrr}).
The scalar curvature $R$ of the metric (\ref{gen}) becomes,
using (\ref{qqq}) and (\ref{rrr}),
\begin{eqnarray}
R&=&\gamma^{\mu\nu}R_{\mu\nu}' + \phi^{a c}R_{a c}
 - {1\over 4}\gamma^{\mu\nu}\gamma^{\alpha\beta}
  \phi_{a b}F_{\mu\alpha} ^ { \  \ a}
  F_{\nu\beta}^{\ \ b}                       \nonumber\\
& &-{1\over 4}\gamma^{\mu\nu}\phi ^ {a b}\phi ^ {c d}\Big\{
  (D_{\mu}\phi_{a c})(D_{\nu}\phi_{b d})
 -(D_{\mu}\phi_{a b})(D_{\nu}\phi_{c d}) \Big\}   \nonumber\\
& &-{1\over 4}\phi ^ {a b}\gamma^{\mu\nu}
  \gamma^{\alpha\beta}\Big\{
 (\partial_{a}\gamma_{\mu \alpha})(\partial_{b}\gamma_{\nu\beta})
 -(\partial_{a}\gamma_{\mu \nu})(\partial_{b}
  \gamma_{\alpha\beta}) \Big\}                        \nonumber\\
& &-(\hat{\nabla}_{\mu}+\hat{\Gamma}_{c \mu}^{\ \ c})j^{\mu}
 -(\hat{\nabla}_{a}
  +\hat{\Gamma}_{\alpha a}^{\ \ \alpha})j^{a}.  \label{to}
\end{eqnarray}
Here $j^{\mu}$ and $j^{a}$ are defined as
\begin{eqnarray}
& &j^{\mu}=\gamma^{\mu\nu}\phi^{a b}\hat{\partial}_{\nu}
\phi_{a b}-2 \gamma^{\mu\nu}
\partial_{a}A_{\nu}^{\ a},       \nonumber\\
& &j^{a}=\phi^{a b}\gamma^{\mu\nu}
  \partial_{b}\gamma_{\mu\nu},
\end{eqnarray}
and $D_{\mu}\phi_{a b}$ is the diff$N_{2}$-covariant derivative
\begin{eqnarray}
D_{\mu}\phi_{a c}&=&\hat{\partial}_{\mu}\phi_{a c}
   - (\partial_{a}A_{\mu}^{\ e})\phi_{e c}
   -(\partial_{c}A_{\mu}^{\ e})\phi_{a e}      \nonumber\\
&=&\partial_{\mu}\phi_{a c}-[A_{\mu}, \phi ]_{a c},
\end{eqnarray}
where $[A_{\mu}, \phi ]_{a c}$ is the Lie derivative of
$\phi_{a c}$ along $A_{\mu}=A_{\mu}^{\ e}\partial_{e}$,
\begin{equation}
[A_{\mu}, \phi ]_{a c} =A_{\mu}^{\ e}\partial_{e}\phi_{a c}
 +(\partial_{a}A_{\mu}^{\ e})\phi_{e c}
 + (\partial_{c}A_{\mu}^{\ e})\phi_{a e}.
\end{equation}
Thus the E-H Lagrangian density in this (2,2)-splitting
finally becomes
\begin{eqnarray}
{\cal L}'&=&\sqrt{-\gamma}\sqrt{\phi} \, R           \nonumber\\
&=& \sqrt{-\gamma}\sqrt{\phi} \, \Big[ \gamma^{\mu\nu}R_{\mu\nu}'
  + \phi^{a c}R_{a c}
  -{1\over 4}\gamma^{\mu\nu}\gamma^{\alpha\beta}\phi_{a b}
   F_{\mu\alpha} ^ { \  \ a}
   F_{\nu\beta}^{\ \ b}                          \nonumber\\
& &-{1\over 4}\gamma^{\mu\nu}\phi ^ {a b}\phi ^ {c d}\Big\{
  (D_{\mu}\phi_{a c})(D_{\nu}\phi_{b d})
 -(D_{\mu}\phi_{a b})(D_{\nu}\phi_{c d}) \Big\}   \nonumber\\
& &-{1\over 4}\phi ^ {a b}\gamma^{\mu\nu}
  \gamma^{\alpha\beta}\Big\{
 (\partial_{a}\gamma_{\mu \alpha})(\partial_{b}\gamma_{\nu\beta})
 -(\partial_{a}\gamma_{\mu \nu})(\partial_{b}
  \gamma_{\alpha\beta})\Big\} \Big]              \nonumber\\
& &-\sqrt{-\gamma}\sqrt{\phi} \,
\Big\{
(\hat{\nabla}_{\mu}+\hat{\Gamma}_{c \mu}^{\ \ c})j^{\mu}
+(\hat{\nabla}_{a}+\hat{\Gamma}_{\alpha a}^{\ \ \alpha})j^{a}
\Big\},                         \label{ac}
\end{eqnarray}
where $\gamma={\rm det} \, \gamma_{\mu\nu}$ and
$\phi={\rm det} \, \phi_{a b}$.
It can be shown that the last two terms in (\ref{ac}) are
total divergences\cite{soh,yoon,yoona},
\begin{eqnarray}
& &\sqrt{-\gamma}\sqrt{\phi} \,
(\hat{\nabla}_{\mu}+\hat{\Gamma}_{c \mu}^{\ \ c})j^{\mu}
 =\partial_{\mu}\Big( \sqrt{-\gamma}\sqrt{\phi} \, j^{\mu} \Big)
 -\partial_{a}\Big( \sqrt{-\gamma}\sqrt{\phi} \,
 A_{\mu}^{\ a}j^{\mu} \Big),                 \label{diver}\\
& &\sqrt{-\gamma}\sqrt{\phi} \,
(\hat{\nabla}_{a}+\hat{\Gamma}_{\alpha a}^{\ \ \alpha})j^{a}
 =\partial_{a}\Big(
  \sqrt{-\gamma}\sqrt{\phi} \,  j^{a}\Big),      \label{diva}
\end{eqnarray}
using (\ref{use}) and the following identities
\begin{eqnarray}
& &\hat{\nabla}_{\mu}j^{\mu}=\hat{\partial}_{\mu}j^{\mu}
  + \hat{\Gamma}_{\alpha \mu}^{\ \ \alpha}j^{\mu}, \nonumber\\
& &\hat{\nabla}_{a}j^{a}=\partial_{a}j^{a}
  + \hat{\Gamma}_{b a}^{\ \ b}j^{a}.
\end{eqnarray}
In the $(u,v)$ coordinates
\begin{equation}
u={1 \over \sqrt{2}}(x^{0} - x^{1}), \hspace{1cm}
v={1 \over \sqrt{2}}(x^{0} + x^{1}),  \label{you}
\end{equation}
the following substitution
\begin{equation}
\gamma_{+ -}=-1, \hspace{.5cm}
\gamma_{+ +}=\gamma_{- -}=0,      \label{your}
\end{equation}
together with
\begin{equation}
A_{\pm}^{\ a}={1\over \sqrt{2}}
(A_{0}^{\ a} \mp A_{1}^{\ a}),  \label{yours}
\end{equation}
leads to the double null gauge (\ref{res}) and
enormously simplifies  the E-H Lagrangian density (\ref{ac}).
The resulting expression is ${\cal L}_{0}$ in (\ref{ga}).

\section{The Covariant Null Tetrads}

In this appendix, we describe the kinematics of a Lorentzian
space-time of 4-dimensions using the covariant null
tetrads\cite{newman,unti,newmantod,pen,hay}. This allows us
to compare the variables in the traditional double null
hypersurface formulation and our KK variables directly.
Moreover, in order to express the volume integral (\ref{ha4})
as a surface integral, it is better to use the covariant null
tetrad notation. Let the two real dual null tetrads
$l_{A}$ and  $n_{A}$ (A=0,1,2,3) be the gradient
fields for some scalar functions $u$ and $v$,
\renewcommand{\theequation}{B\arabic{equation}}
\setcounter{equation}{0}
\begin{equation}
l_{A}=\nabla_{A}u, \hspace{2cm}
n_{A}=\nabla_{A}v,
\end{equation}
so that
$\nabla_{{[}B}l_{A{]}}=\nabla_{{[}B}n_{A{]}}=0$.
The dual vector fields $du$ and $dv$ are related to the
dual null tetrads by
\begin{equation}
du=l_{A}dX^{A}, \hspace{2cm}
dv=n_{A}dX^{A}.
\end{equation}
We also have the vector fields
${\partial / \!  \partial u}$, ${\partial / \! \partial v}$, and
${\partial / \!  \partial y^{a}}$ $(a=2,3)$ which we may write
\begin{equation}
{\partial \over \partial u}=u^{A}
{\partial \over \partial X^{A}}, \hspace{.5cm}
{\partial \over \partial v}=v^{A}
{\partial \over \partial X^{A}}, \hspace{.5cm}
{\partial \over \partial y^{a}}=y_{a}^{\ A}
{\partial \over \partial X^{A}}.
\end{equation}
If we choose the basis vector fields of space-time so that
\begin{math}
\nabla_{A}=(\partial / \!  \partial u, \partial / \!  \partial v,
\partial / \!  \partial y^{a}),
\end{math}
the components of the dual null tetrads and vector
fields are given by
\begin{eqnarray}
& &l_{A}=(1,0,0,0) \hspace{1cm} n_{A}=(0,1,0,0),  \nonumber\\
& &u^{A}=(1,0,0,0),\hspace{1cm} v^{A}=(0,1,0,0),  \hspace{1cm}
   y_{a}^{\ A}=\delta_{a} ^{\ A}.                 \label{comp}
\end{eqnarray}
{}From this it follows that
\begin{equation}
l_{A}u^{A}= n_{A}v^{A}=1, \hspace{2cm}
l_{A}v^{A}=n_{A}u^{A}=0.                        \label{rel}
\end{equation}
Since $l_{A}l^{A}=n_{A}n^{A}=0$, the null tetrad $l^{A}$
and $n^{A}$ may be chosen as
\begin{equation}
l=l^{A}{\partial\over \partial X^{A}}=-{\partial\over \partial v}
  + A_{-}^{\ a}{\partial\over \partial y^{a}}, \hspace{.5cm}
n=n^{A}{\partial\over \partial X^{A}}=-{\partial\over \partial u}
  + A_{+}^{\ a}{\partial\over \partial y^{a}},
\end{equation}
i.e.
\begin{equation}
l^{A}=(0,-1, A_{-}^{\ a}), \hspace{.5cm}
n^{A}=(-1,0, A_{+}^{\ a}),
\end{equation}
such that
\begin{equation}
l_{A}n^{A}=-1.                              \label{norm}
\end{equation}
($n$ and $l$ are the minus of the horizontal lift vector fields
\begin{math}
\hat{\partial}_{\pm}=\partial_{\pm}-A_{\pm}^{\ a}\partial_{a}
\end{math}
in the $(u,v)$-coordinates, respectively.)
The condition (\ref{norm}) is equivalent to the previous
normalization condition $g^{+-}=-1$ in (\ref{inv}),
and means that, given
an arbitrary function $u$, the function $v$ must be chosen
in such a way that the normalization condition (\ref{norm})
is satisfied. We still have the freedom to orient $l^{A}$
and $n^{A}$ in space-time, and we fix this freedom
by demanding that $l^{A}$ and $n^{A}$ are normal to the
2-dimensional spacelike surface $N_{2}$ whose tangent
vector fields are $\partial / \!  \partial y^{a}$,
\begin{equation}
h_{A B}l^{B}=h_{A B}n^{B}=0,              \label{orth}
\end{equation}
where $h_{A B}$ is the metric on $N_{2}$ ($h_{A B}$ is
the covariant expression of $\phi_{a b}$).
Using $h_{A B}$ and
$l_{A}$, $n_{A}$, the space-time metric $g_{A B}$ may be
written as
\begin{equation}
g_{A B}=h_{A B} - (l_{A}n_{B} + n_{A}l_{B}).     \label{mett}
\end{equation}
{}From these relations, we easily find that
(\ref{mett}) is identical to the metric
(\ref{res})
\begin{equation}
ds^{2} =- 2 du dv + \phi_{a b} (A_{+}^{\ a} du
       + A_{-}^{\ a} dv + dy^{a})
       (A_{+}^{\ b} du +A_{-}^{\ b} dv + dy^{b}).  \label{metta}
\end{equation}

\section{The Bondi Surface Integral}

We now show that the volume integral (\ref{ha4})
\renewcommand{\theequation}{C\arabic{equation}}
\setcounter{equation}{0}
\begin{equation}
E={1\over 3}\int dv d^{2}y \ {\rm e}^{\sigma}
h^{A C}h^{B D} C_{A B C D}            \label{ha11}
\end{equation}
can be expressed as a surface integral,
using the Bianchi identity of the conformal curvature tensor
\begin{equation}
\nabla_{{[}M}C_{A B{]} C D}=0.              \nonumber
\end{equation}
Let us notice that, due to the Bianchi identity,
the following is true for any scalar
function $\tilde{\Omega}$,
\begin{equation}
\nabla_{{[}M}\Big(\tilde{\Omega}^{-1}C_{A B{]} C D}\Big)=
\Big(\nabla_{{[}M}
\tilde{\Omega}^{-1}\Big)C_{A B{]} C D}.  \label{eq}
\end{equation}
If we contract (\ref{eq}) by $h^{A C}h^{B D}$, it becomes
\begin{equation}
h^{A C}h^{B D}
\nabla_{{[}M}\Big(\tilde{\Omega}^{-1}C_{A B{]} C D}\Big)
=h^{A C}h^{B D}\Big(
\nabla_{{[}M}\tilde{\Omega}^{-1}\Big)C_{A B{]} C D}. \label{eq1}
\end{equation}
Let us choose $\tilde{\Omega}^{-1}$ as a function of $(u,v)$ only,
and let $\nabla_{M}=\nabla_{-}$. Then the r.h.s. of
(\ref{eq1}) becomes
\begin{equation}
h^{A C}h^{B D}\Big(
\nabla_{{[}-}\tilde{\Omega}^{-1}\Big)C_{A B{]} C D}
={1\over 3}h^{A C}h^{B D}
\Big(\nabla_{-}\tilde{\Omega}^{-1}\Big)C_{A B C D}, \label{rhs}
\end{equation}
since $h^{A B}\nabla_{B}\tilde{\Omega}^{-1}=0$.
The l.h.s. of (\ref{eq1}) becomes
\begin{eqnarray}
h^{A C}h^{B D}
\nabla_{{[}-}\Big(\tilde{\Omega}^{-1}C_{A B{]} C D}\Big)
&=&\nabla_{{[}-}\Big(
h^{A C}h^{B D}\tilde{\Omega}^{-1}C_{A B{]} C D}\Big)   \nonumber\\
& &- \nabla_{{[}-}\Big( h^{A C}h^{B D} \Big)
\tilde{\Omega}^{-1}C_{A B{]} C D}.           \label{eq2}
\end{eqnarray}
Using $h^{A C}=g^{A C}+l^{A}n^{C} + n^{A}l^{C}$
and the properties of the conformal curvature tensor
\begin{eqnarray}
& &g^{A C}C_{A B C D}=g^{B D}C_{A B C D}=0, \nonumber\\
& &C_{A B C D}=C_{[A B][C D]},
\end{eqnarray}
the second term in the r.h.s. of (\ref{eq2}) may be written as
\begin{equation}
\nabla_{{[}-}
\Big( h^{A C}h^{B D}\Big) \tilde{\Omega}^{-1}C_{A B{]} C D}
=-2\nabla_{{[}-} \Big(l_{A}n_{B{]}}l_{C}n_{D}\Big)
\tilde{\Omega}^{-1}C^{A B C D}.                     \label{eq5}
\end{equation}
Since $l_{A}$ and $n_{A}$ are non-zero only when
$A,B$ are $+$ or $-$, we have
\begin{equation}
\nabla_{{[}-} \Big(l_{A}n_{B{]}}l_{C}n_{D}\Big)=0,  \label{haha}
\end{equation}
due to the repeated indices in the anti-symmetric symbol.
Therefore the l.h.s. of (\ref{eq1}) becomes
\begin{equation}
h^{A C}h^{B D}
\nabla_{{[}-}\Big(\tilde{\Omega}^{-1}C_{A B{]} C D}\Big)
=\nabla_{{[}-}\Big(
h^{A C}h^{B D}\tilde{\Omega}^{-1}C_{A B{]} C D}\Big). \label{oka}
\end{equation}
Thus (\ref{eq1}) becomes, using (\ref{rhs})
and (\ref{oka}),
\begin{eqnarray}
h^{A C}h^{B D}\Big(
\nabla_{-}\tilde{\Omega}^{-1}\Big) C_{A B C D}&=&
\nabla_{-}\Big( h^{A C}h^{B D}\tilde{\Omega}^{-1}C_{A B C D}\Big)
+\nabla_{A}\Big(
h^{A C}h^{B D}\tilde{\Omega}^{-1}C_{B - C D}\Big)    \nonumber\\
& &+\nabla_{B}\Big(
h^{A C}h^{B D}\tilde{\Omega}^{-1}C_{- A C D}\Big). \label{ok}
\end{eqnarray}
Integrating (\ref{ok}) over the $u={\rm constant}$
hypersurface with the canonical measure $\sqrt{h}$\cite{wald},
it becomes, using $\sqrt{g}=\sqrt{h}$ for the metric
(\ref{mett}),
\begin{eqnarray}
\int dv d^{2}y \sqrt{h} \,  h^{A C}h^{B D}
   \Big(\nabla_{-}\tilde{\Omega}^{-1}\Big)C_{A B C D}
&=&\int dv d^{2}y\sqrt{h} \, \nabla_{-}
\Big( h^{A C}h^{B D}
\tilde{\Omega}^{-1}C_{A B C D}\Big)   \nonumber\\
&=&\lim \int d^{2}y \sqrt{h} \,
 \Big( h^{A C}h^{B D}
\tilde{\Omega}^{-1}C_{A B C D}\Big),       \label{ano}
\end{eqnarray}
since the surface integrals coming from the last two terms in
the r.h.s. of (\ref{ok}) vanish for any 2-surface $N_{2}$
compact without boundary. Here
$\lim$ means that the integral over $N_{2}$ must
be evaluated at the limiting boundary value(s) of $v$.
Now let us choose $\tilde{\Omega}$ such
that $\tilde{\Omega}^{-1}=v$. Then the identity
(\ref{ano}) becomes
\begin{equation}
\int dv d^{2}y \sqrt{h} \,  h^{A C}h^{B D}C_{A B C D}
=\lim \ v \!\! \int d^{2}y \sqrt{h} \,
h^{A C}h^{B D}C_{A B C D}.               \label{anoth}
\end{equation}
Since $\sqrt{h}={\rm e}^{\sigma}$ in our previous notation,
the volume integral (\ref{ha4}) becomes
\begin{equation}
E={1\over 3}\int dv d^{2}y \ {\rm e}^{\sigma}
   h^{A C}h^{B D}C_{A B C D}
 ={1\over 3}\lim \ v \!\! \int d^{2}y \ {\rm e}^{\sigma}
   h^{A C}h^{B D}C_{A B C D},       \label{anothe}
\end{equation}
which is the desired expression for the Bondi surface integral as
$v$ approaches to infinity.

\section{The Positive-Definiteness of $\kappa_{\pm}^{2}$}

Here we show that $\kappa_{\pm}^{2}$ is positive-definite.
Let us introduce super indices $A',B'$ for the symmetric
combinations $(a c)$ and $(b d)$, respectively, so that
\renewcommand{\theequation}{D\arabic{equation}}
\setcounter{equation}{0}
\begin{equation}
\rho_{A'}:=\rho_{a c}, \hspace{1cm}
\rho_{B'}:=\rho_{b d}.
\end{equation}
Define the supermetric $G_{A' B'}$  and its inverse $G^{A' B'}$
by
\begin{equation}
G_{A' B'}:={1\over 2}(\rho_{a b}\rho_{c d}+\rho_{a d}\rho_{c b}),
\hspace{1cm}
G^{A' B'}:={1\over 2}(\rho^{a b}\rho^{c d}+\rho^{a d}\rho^{c b}),
\end{equation}
such that
\begin{equation}
G^{A' E'}G_{E' B'}=\delta^{A'}_{\ B'}, \hspace{.5cm}
{\rm where}\ \ \ \ \
\delta^{A'}_{\ B'}={1\over 2}(\delta^{a}_{\ b}\delta^{c}_{\ d}
 + \delta^{a}_{\ d}\delta^{c}_{\ b}).
\end{equation}
This supermetric raises and lowers the super indices
\begin{equation}
G^{A' B'}\rho_{B'}=\rho^{A'}, \hspace{1cm}
G_{A' B'}\rho^{B'}=\rho_{A'},
\end{equation}
and has a positive-definite signature
since it becomes
\begin{equation}
G^{A' B'}={\rm diag}(+1, +1/  2, +1)\hspace{.5cm}
{\rm for}\ \ \ \rho^{a b}=\delta^{a b}.
\end{equation}
Therefore it follows that
\begin{eqnarray}
\kappa_{\pm}^{2}&=&{1\over 8}\rho^{a b}\rho^{c d}
(D_{\pm}\rho_{a c}) (D_{\pm}\rho_{b d})       \nonumber\\
&=&{1\over 8}G^{A' B'}(D_{\pm}\rho_{A'})(D_{\pm}\rho_{B'})
\geq 0.
\end{eqnarray}

\end{document}